\documentclass[12pt]{iopart}
\usepackage[colorlinks]{hyperref}
\usepackage{graphicx}
\usepackage{url}
\usepackage[titletoc]{appendix}
\usepackage{tcolorbox}
\usepackage [latin1]{inputenc}
\begin{document}
\hypersetup{
  colorlinks   = true, 
  urlcolor     = blue, 
  linkcolor    = blue, 
  citecolor   = blue 
}
\thispagestyle{empty}
\begin{center}
{\Large\bf Electromagnetic induction: physics, historical breakthroughs, epistemological issues  and textbooks}\\
\vskip4mm
Giuseppe Giuliani\\
\vskip4mm
\small Formerly,  Dipartimento di Fisica, Universit\`a degli Studi, via Bassi 6, 27100 Pavia, Italy. 
\vskip4mm
giuseppe.giuliani@unipv.it
\end{center}
\vskip5mm
{\bf Abstract.} The discovery of Electromagnetism by {\O}rsted (1820) initiated an ``extraordinary decennium'' ended by the discovery of electromagnetic induction by Faraday (1831). During this decennium, in several experiments, the electromagnetic induction was there, but it was not seen or recognized.
\par\noindent
 In 1873, James Clerk Maxwell, within a Lagrangian description of electric currents, wrote down a `general law of electromagnetic induction' given by, in modern form and with standard symbols:
$$
     \mathcal E = \oint_l [\vec E+(\vec v\times \vec B)]\cdot\vec{dl}; \qquad \vec E = -\nabla\varphi -\frac{\partial \vec A}{\partial t}
$$
In Maxwell's derivation, the velocity appearing in this equation, is the velocity of the line element $\vec {dl}$.
A modern reformulation of Maxwell's general law, starts with the definition of the induced {\em emf} as the $\mathcal E= \oint_l [\vec E+(\vec v_c\times \vec B)]\cdot\vec{dl}$, where $\vec v_c$ is the velocity of the charge. It is shown that, this apparently minor difference, is fundamental.
The general law is a local law: it correlates what happens in the line at the instant $t$ to the values of quantities at the points of the line at the same instant $t$. For rigid circuits, it is Lorentz invariant. If expressed in terms of the magnetic field, it allows -- in the approximation of low velocities -- the derivation of the ``flux rule'', for {\em filiform} circuits. The ``flux rule'' is a calculation tool and not a physical law because  it not always yields the correct prediction, it does not say where the induced {\em emf} is localized, it requires {\em ad hoc} choices of the integration paths and   -- last but not least -- because, if physically interpreted,  it implies physical interactions with speeds higher than that of light.
\par\noindent
Maxwell's general law has been rapidly forgotten; instead, the ``flux rule'' has deeply taken root.
 The reasons appear to be various. One is plausibly related to the idea that the vector potential does not have physical meaning, a stand clearly assumed by Hertz and Heaviside at the end of the Nineteenth century. An analysis of university textbooks, spanned over one century, assumed to be representative on the basis of the authors and/or on their popularity or diffusion,  suggests that other reasons  are related to the habit of presenting electromagnetic phenomena following the historical development and to the epistemological stand according to which physical laws must be derived from experiment without the need of recurring, soon or later, to an axiomatic presentation of the matter.
 On the other hand, the rooting of the ``flux rule'' has been certainly favored by its calculation utility: this practical feature has overshadowed its predictive and epistemological weakness.
 \par\noindent
 In the first decades of the Twentieth century,  it was  common the idea that some electromagnetic induction experiments with rotating cylindrical magnets could be explained also by assuming that the ``lines of magnetic force'' introduced by Faraday  rotate with the magnet. This is a surprising hypothesis, if one takes into account the fact that Faraday's experiments, as repeatedly stressed by him, prove that the ``lines od magnetic force'' do not rotate. More surprisingly, this hypothesis has been resumed recently. It is shown that the hypothesis of rotating ``lines of magnetic force'' is incompatible with Maxwell - Lorentz - Einstein electromagnetism and is falsified by experiment. Finally, the electromagnetic induction in some recent research papers is briefly discussed.

\vskip4mm
\section{Introduction}
From the point of view of electromagnetic phenomena, the Nineteenth century can be characterized by two events separated by almost one hundred years: the invention of the pile by Alessandro Volta (1800) and the discovery of the electron by Joseph John Thomson (1897).
Volta's invention has been a cornerstone in the technological evolution and furnished scientists an unthought  tool for generating (nearly) continuous electrical currents, thus allowing the investigation of their properties. Thomson's discovery showed the discreteness  and material (endowed with mass) nature of the electric charge.
In between, there has been an intricate maze of experiments, models, theories, philosophical  and epistemological positions. But, at the end, the physicists  found a way out. They unified the electric and magnetic phenomena, passed from an action at a distance view to a field centered one and progressively acquired a discrete description of matter and electricity, so preparing the abandon of the ether.
\par
Electromagnetic induction constitutes a fundamental chapter of this history.
It was discovered in 1831 by Michael Faraday (1791 - 1867), at the end of an ``extraordinary decennium'' begun with {\O}rsted's discovery of the effect of an electric current on a magnetic needle (1820) (section \ref{pro}).  During this decennium, in several experiments, the electromagnetic induction was there, but it was not seen or recognized (section \ref{unseen}). Over more than twenty five  years, Faraday tried to establish a coherent description of the induction phenomena, based on the idea that there is an induced current when there is an intersection between  a conductor and  the ``lines of magnetic force'' in relative motion. Though not developed in mathematical form, this description can be considered as a theory, capable also of  quantitative accounts  (section \ref{discovery}). In the following years, many physicists tackled the problem both on the experimental and the theoretical side.
On the experimental side, it was not so easy to add novel knowledge to what Faraday had already discovered. An exception was the rule discovered by Heinrich Friedrich Emil Lenz (1804 - 1865) that the induced current is such as to  oppose the phenomenon that generated it \cite{lenz}.
Though qualitative, this rule was conceptually relevant and guided the following theoretical attempts to build up a theory of electromagnetic induction starting from experimental data. Among these attempts,  those by Franz Neumann (1798 - 1895)  and  Wilhelm Weber (1804 - 1891) \cite[pp. 222 - 232]{edmund}, \cite[chapter 2]{darrigol}. However, we must wait for Maxwell's {\em Treatise} for finding a {\em general  law} of electromagnetic induction, {\em derived} within a Lagrangian treatment of the currents (section \ref{maxwell}).
The astonishing feature of this law is that it was obtained without knowing what an electrical current is, apart from the recognition that it is a ``kinetic process''.
This limited knowledge is reflected by the fact that Maxwell's derivation considers the velocity of an elementary circuit element instead of the velocity of the charges contained in it. However,  the  definition of the induced {\em emf} as:
 \begin{equation}\label{femintro}
    \mathcal E = \oint_l (\vec E + \vec v_c \times \vec B)\cdot \vec{dl}; \qquad \vec E=- \nabla\varphi-\frac{\partial\vec A}{\partial t}
 \end{equation}
  where $\vec v_c$ is the velocity of the charge, leads straightforwardly to Maxwell's law with the correct velocity in it (section \ref{mle}).
   Maxwell's general law   has been rapidly forgotten; meanwhile, the ``flux rule'' took root, in spite of the fact that it is just a calculation tool and not a physical law (section \ref{regola}). In section \ref{why}, we try to understand the reasons for such an oblivion, by taking into account also the role played by textbooks.
  \par
   In the first decades of the Twentieth century,  it was  common the idea that some electromagnetic induction experiments with rotating cylindrical magnets could be explained also by assuming that the ``lines of magnetic force'' introduced by Faraday  rotate with the magnet. This is a surprising hypothesis, if one takes into account the fact that Faraday's experiments, as repeatedly stressed by him, prove that the ``lines od magnetic force'' do not rotate. More surprisingly, this hypothesis has been resumed recently. It is shown that the hypothesis of rotating ``lines of magnetic force'' is incompatible with standard electromagnetism and that it is falsified also by recent experiments.
  Finally, the electromagnetic induction in some recent papers is briefly discussed (section \ref{recent}).
  \par
  The treatment of this matter is preceded by two other sections. The first one is dedicated to the names used in Electromagnetism and to the shifts of their meanings (section \ref{names}). The other, to the meaning of the locution `physical meaning', frequently  used in physical literature (section \ref{meaning}).
\par
The discussion on the explanation of electromagnetic induction phenomena has been periodically reopened in spite of the fact that the experimental foundations are well established. This constitutes a rare case in the history of Physics. This paper tries to understand why.

    \section{The names and the things}\label{names}
Along the Nineteenth century, new names have been invented to denote new phenomena or new theoretical entities or  to newly label  already known phenomena. The knowledge inherited from the past centuries about  electricity and magnetism, has been revolutionized by {\O}rsted's discovery of the action of a current carrying wire on a magnetic needle. Electricity and magnetism manifested themselves as intertwined; hence the new term {\em Electromagnetism}, which shined  in the title of an {\O}rsted's paper  \cite{oersted21}. Soon after {\O}rsted's discovery, Amp\`ere coined the term {\em Electrodynamics} for describing the interactions between currents, between currents and magnets and between magnets. Consequently, the  interaction between electrically charged bodies or between magnetic poles at rest, described by Coulomb, was denoted as {\em Electrostatics} and {\em Magnetostatics}. The terms {\em Electromagnetism} and {\em Electrodynamics} appear usually in the titles of our textbooks with, maybe, the specification of `modern'.
\par
Faraday, after the discovery of electromagnetic induction, speaks of  ``volta - electric'' induction and of ``magnetic - electric'' induction depending on whether the inducing agent is a current carrying circuit  or a magnet. The term ``electromagnetic induction'' comes in later.
Faraday's theory of electromagnetic induction is based on the concept of ``lines of magnetic force'', a concept still in use today for describing graphically the properties of the magnetic field. The lexical renewal is a slow process, often  accompanied by shifts in meaning. Thus Faraday's {\em Experimental Researches} are {\em in Electricity} and Maxwell's Treatise is titled {\em A Treatise on Electricity and Magnetism}.
Maxwell denoted by the term ``electric field'' ``the portion of space near electrified bodies, considered with reference to
electrical phenomena'' \cite[p. 45] {treatise1}; the term ``magnetic field''  had also a similar meaning.
Our electric field $\vec E$ is denoted by Maxwell as ``electromotive intensity'', and the force exerted by $\vec E$ on a charged body  is called ``electromotive force'' \cite[p. 46]{treatise1}. The line integral $\int_l \vec E\cdot \vec {dl}$ is called ``total electromotive force''. There is some ambiguity here, because the adjective `total' is often omitted and,
 throughout  the {\em Treatise},  the term ``electromotive force'' is used to denote both  the ``electric field '' and its line integral.
\par
In the manuals of the Twentieth century and in the contemporaries ones, it is found that the magnetic field  $\vec B $, recovering a Nineteenth  century's denotation, is called ``magnetic induction vector'' and the field $ \vec H $ whose sources are the current densities $ \vec J $  is called  ``magnetic field''. To then have to stress that what appears in the expression of the Lorentz force is the magnetic induction vector and not the magnetic field. Not to mention the conceptual  confusion created by the attempt to recall contributions by different researchers in the denotation of a formula, often distorting the history. Then we read of the ``Faraday - Neumann - Lenz law'', or variants at will. The presence of Lenz is justified by the sign (-) that appears in the formula; but Faraday, with the ``flux rule'' has nothing to do.
In fact, as we shall see (section \ref{fartheory}), Faraday developed a {\em local} theory of electromagnetic induction, based on the idea that there is an induced current if there is an intersection between the `lines of magnetic force' and a conductor in relative motion.
The same is true for Franz Neumann (1798 - 1895) whose contribution to the understanding of electromagnetic induction consisted of a formula obtained in the framework of Amp\`ere Electrodynamics, within an action at a distance point of view. It is true that, with hindsight,  one can pick up the vector potential in Neumann's formula; however, Neumann did not identify it as such.
This reminds us that the names of the things  take root in spite of the shift of their meanings. This phenomenon presents two risks.  The conservation of old names in a changed conceptual framework may  induce confusion; conversely, to neglect the shift in meaning of a name, might lead, retrospectively, to wrongly attribute a law or the first use of a concept.
 Not to neglect the Lorentz force. Often, only the component due to the magnetic field is denoted by this name; or, sometimes, we speak of  the ``Coulomb - Lorentz law''.
And, considering the subject of this paper, we cannot fail to conclude with Electromagnetism.  If it is true that Maxwell's equations we use today are the same as Maxwell's (though written in another form), we can not  forget that the Nineteenth century  interpretations of Maxwell's equations are no longer our's. In an axiomatic presentation of Maxwell's equations, the physical dimensions of the fields is given by the Lorentz force, while  the inherently relativistic nature of the theory has been highlighted by Einstein. Therefore, not to recall various contributions in one name, but to underline our interpretation of the theory, we should use the term of Maxwell - Lorentz - Einstein (MLE) Electromagnetism. This term will be used from now on.
\section{The meaning of `physical meaning'}\label{meaning}
In physical literature, we often encounter the expression `physical meaning', generally used without specifying its meaning. In these cases, the meaning of `physical meaning' is suggested by the context. A `physical meaning' can be attributed to
different things: theoretical entities, physical quantities, concepts, single statements, sets
of statements.
Here, we propose to use the expression in a technical way, based on clear  definitions.
Let us begin by briefly recalling that
examples of theoretical entities are: material point, rigid body, gas, electron. Instead, physical quantities  describe properties of theoretical entities, or are attributed to theoretical entities, or describe interactions between theoretical entities. For example: the mass describes a property of a material point, a rigid body, a gas or an electron; the velocity, with respect to a reference system, is attributed to a material point, a rigid body or an electron; the force describes the interaction between two rigid bodies or between two electrons.
\par
Some theoretical entities  used in  the past  are no more in use; for example: caloric, ether.
It is reasonable to think that, in  those times, these theoretical entities had
physical meaning. Therefore,   we should  find out  criteria  that are, at the same time, prescriptive but
historically flexible. The following seem to satisfy these conditions:
\begin{itemize}
  \item  A theoretical entity has physical meaning  if its
       elimination reduces the predictive power of a theory (strong condition).
         \item And/Or: a theoretical entity has physical meaning  if its
       elimination reduces the descriptive or the heuristic capacity of a theory (weak condition).
\end{itemize}
This criterion can be considered as a variant of the one enunciated by Hertz:
\begin{quote}\small
I have further endeavoured in the exposition to limit as far as possible the number of those conceptions which are arbitrarily introduced by us, and only to admit such elements
as cannot be removed or altered without at the same time altering possible experimental results \cite[p. 28]{ew}.
\end{quote}
A significant example for the application of these conditions is
provided by the concept of ether. In Maxwell's times, the
electromagnetic field was conceived as a mechanical deformation of the ether.
 Maxwell's theory apart, light was conceived as a
wave phenomenon that develops in an ether. So, the ether had  physical meaning  because its elimination would have reduced the predictive power of those theories.
 For the weak condition,
 we can consider the concept of  space - time.
Special relativity  was born and can still be presented axiomatically
without the use of the space - time formalism. However, it is evident that the use
of this formalism increases the descriptive capacity of the theory.
On the other hand, it can not be denied
the heuristic role played by the space - time concept  in the birth
of general relativity.
\par
For  physical quantities, a similar criterion is the following. A physical quantity has physical meaning if:
\begin{itemize}
  \item   \label{pq}     Its
       elimination reduces the predictive power of a theory (strong condition).
        \item And/or  if its
       elimination reduces the descriptive capacity of a theory (weak condition).
  \end{itemize}
   In general, a physical quantity can be measured, at least in principle. An interpreted theory indicates or suggests which physical quantities are to be measured.
Usually, this  feature is considered as a sufficient condition for attributing physical meaning to a physical quantity. However, it is easy to see that this is not the case. For example, in the item {\em Ether} written for the ninth edition of the {\em Enciclop{\ae}dia Britannica}, Maxwell showed how to measure  the rigidity of the ether by measuring the intensity of the Sun's light impinging on the Earth \cite{etere}. However, the physical meaning of the ether's rigidity is not assured by the fact that it can be measured (within Maxwell's interpretation of his theory), but by the first condition above. The above criterion connects intimately the physical meaning of a physical quantity to the theories that use it. So, the ether and its rigidity had physical meaning within Maxwell's theory; but they have lost any meaning in MLE electromagnetism.
\par
In the light of the above criteria, it is worth considering the case of the wave - function. It is a physical quantity that can describe -- in principle -- the properties of every physical system. If one finds the expression of the wave - function of, say, the electron of the Hydrogen atom, then the wave - function allows to predict the expectation values of some specific physical quantities of the electron. However, the wave - function  can not be measured as such. For this reason, the possibility of being measured can not be considered as a necessary condition for attributing physical meaning to a physical quantity.
\par
We shall use the above criteria in due time.
 \section{The discovery of Electromagnetism}\label{pro}
At the onset of the Nineteenth century, the idea that electricity and magnetism could be in some way connected was in the air. In 1805, Jean Nicolas Pierre Hachette (1769 - 1834) and Charles Bernard Desormes (1777 - 1862) after having seen a magnet bar  placed on a wooden boat floating on quiet water  to align  along the magnetic meridian, looked for a similar behavior of a large voltaic pile \cite{hachette}. But they did not observed any rotation. Hachette recalled this experiment also in his (September, 1820) report on the discoveries by {\O}rsted and Amp\`ere \cite{hachette2}. This experiment  appears to us  very naive; however, it must be taken into account that the concept of electrical circuit and that of electrical current as electricity in motion  began to emerge only  after Amp\`ere's interpretation of {\O}rsted's experiment \cite{francesi}. This situation is unambiguously reflected by the fact that our {\em conducting} wire connecting the two poles of a battery, was systematically  denoted as the {\em connecting} wire. The static conception of electricity was still undiscussed.
 \par
The turning point was due to Hans Christian {\O}rsted (1777 - 1851). His discovery (July, 1820) of the deflection of a magnetic needle by a current carrying wire opened the way towards the unification of electric and magnetic phenomena \cite{oersted}. {\O}rsted's discovery has been by no means accidental. He was convinced of a connection between electricity and magnetism, within an unitary view of the chemical, thermal, electrical and magnetic forces  \cite{stauffer1, stauffer2}.
The idea that {\O}rsted's discovery was accidental has been suggested by Ludwig Wilhelm Gilbert (1769 - 1824), then editor of the {\em Annalen der Physik}, in the presentation of the German translation of {\O}rsted's Latin text.
  With these words: ``What all research and effort had not produced, happened by chance to Professor {\O}rsted in Copenhagen, during his lectures on electricity and magnetism last winter'' \cite[p. 292]{gilbert}.
  This claim was groundless: in fact, {\O}rsted had provided no indication as to how the initial observation had been made.
  This drastic position has been traced  back   to Gilbert's long cultural battle against a speculative approach to science, typical of the supporters of {\em Naturphilosophie}, among whom, evidently, was considered  also {\O}rsted \cite{stauffer1, stauffer2, gerard}.
  \par
An year after his discovery, {\O}rsted, in a work entitled {\em Considerations on Electromagnetism}, tries to disprove Gilbert's version and dedicates, to this aim, the first section entitled {\em To serve the history of my previous works on this topic} \cite{oersted21}.
In particular, {\O}rsted recalls that, in 1812, he had written: ``One should try to see if electricity in its most latent state can produce some effect on a magnet''. However, between this very generic indication and the conception and realization of the experiment, about eight years had passed by.
     Indeed, {\O}rsted admits that: ``I wrote this, during a journey;
therefore I could not easily start the experiments;
{\em on the other hand, the way of doing them was not at all
clear to me at that moment}, all my attention being focused on the development
of a  chemistry system \cite[p. 162, french transl.; italics mine]{oersted21}.''
And, recalling how the first experiment took place during the lesson, {\O}rsted writes:
\begin{quote} \small
My old belief about the unity of electric and magnetic forces had developed with new clarity and I decided to submit my opinion to  experiment. Preparations were made one day when I was supposed to give a lesson that same evening. Then, I showed  Canton's experiment on the influence of chemical effects on the magnetic state of iron; I called attention to the changes in the magnetized needle during a storm and {\em conjectured that an electric discharge might have some effect on a magnetic needle placed outside the galvanic circuit}. I decided to do the experiment right away. Since I was expecting the maximum [effect] from a discharge that produces incandescence, I inserted a very thin platinum wire at the position of the needle, placed under [the wire]. Although the effect has been indisputable, it nevertheless seemed so confusing to me that I decided to investigate the matter further when I had more time available \cite[p. 163, french transl.; italics mine] {oersted21}.
\end{quote}
Although {\O}rsted gives some  details about the apparatus used - the glowing platinum wire - he gives no indication as to what he actually observed. The last sentence adds, if possible, more ambiguity: the effect was ``indisputable'' but ``confusing''.
\par
 A recollection by Christopher Hansteen (1784 - 1873) containing rather interesting details, absent from {\O}rsted's account, appeared in 1870. It brought water to the mill of an accidental discovery.
   In 1857, that is almost forty years after the events narrated, Hansteen wrote a letter to Faraday. This was published in the volume {\em Life and Letters of Faraday} by Henry Bence Jones (1813 - 1873) \cite[p. 390]{jones}.
  In the letter, Hansteen describes {\O}rsted as an ``unhappy experimenter'', always in need of help;
   he then describes a lesson during which {\O}rsted would have performed the famous experiment without success.
   The whole passage  is worth reading:
  \begin{quote}\small
  Professor {\O}rsted was a man of genius, but he was a
very unhappy experimenter; he could not manipulate
instruments. He must always have an assistant, or one
of his auditors who had easy hands, to arrange the
experiment; I have often in this way assisted him as
his auditor. Already in the former century there was
a general thought that there was a great conformity,
and perhaps identity, between the electrical and magnetical
force; it was only the question how to demonstrate
it by experiments. {\O}rsted tried to place the
wire of his galvanic battery perpendicular (at right
angles) over the magnetic needle, but remarked no
sensible motion. Once, after the end of his lecture, as
as he had used a strong galvanic battery to other
experiments, he said, ``Let us now once, as the battery
is in activity, try to place the wire parallel with the
needle;" as this was made, he was quite struck with
perplexity by seeing the needle making a great oscillation
(almost at right angles with the magnetic meridian).
Then he said, ``Let us now invert the direction of the
current," and the needle deviated in the contrary
direction. Thus the great detection was made; and it
has been said, not without reason, that ``he tumbled over it by accident." He had not before any more
idea than any other person that the force should be
{\em transversal}. But as Lagrange had said of Newton in
a similar occasion, ``such accidents only meet persons
who deserve them \cite[pp. 390 - 391]{jones}."
  \end{quote}
For a historical reconstruction of this testimony see, for example, \cite {stauffer1, kipnis}. In particular, Stauffer argues that Hansteen's ``testimony'' is not direct because, at the time of the events narrated, he was not in Copenhagen \cite [p. 309] {stauffer1}; while Kipnis describes in greater detail the changes, over time, in the appreciation of {\O}rsted's work \cite[pp. 3 - 4]{kipnis}.
  The motivation that led Hansteen to include this recollection in a rather generic letter addressed to Faraday is not clear. From the text, no motivation emerges for telling this episode to Faraday. However, the sentence following the text quoted above reads:
\begin{quote}\small
    You completed the detection by inverting the experiment
by demonstrating that an electrical current can
be excited by a magnet, and this was no accident, but
a consequence of a clear idea. I pretermit your many
later important detections, which will conserve your
name with golden letters in the history of magnetism \cite[p. 391]{jones}.
\end{quote}
This sentence, contrasting {\O}rsted's ``casual'' discovery with that rationally sought by Faraday, suggests that Hansteen's aim was to earn Faraday's benevolence. Whatever the case,
   let us assume that Hansteen's account is a faithful report of what happened.
     That {\O}rsted was a clumsy experimenter is a typical {\em ad hominem} argument, irrelevant to any evaluation of what {\O}rsted discovered. The testimony that, initially, {\O}rsted would have placed the wire perpendicular to the direction of the magnetic needle appears plausible for reasons of symmetry. In fact, a correct arrangement - with the wire {\em parallel} to the direction of the needle (oriented along the magnetic meridian) - would have implied, if  we are allowed to use an anachronistic term, a  symmetry breaking: why the needle should choose between East and West, a priori equivalent, in the absence of a hardly conceivable hypothesis of an {\em asymmetrical} action on the needle by the wire? In the light of these reflections, {\O}rsted's reaction of astonishment at the discovery made with the wire parallel to the needle becomes plausible. If things have unfolded as Hansteen recounts, {\O}rsted's attempt, at the end of the lesson, to try with wire and needle parallel only demonstrates the experimenter's tenacity to prove the existence of a long - hypothesized phenomenon. If rationally interpreted, Hansteen's recollection turns into a ``testimony '' in favor of the thesis that it was not an accidental discovery.
     It is therefore not clear why Hansteen, at the end  inserts the sentence ``Thus was made the great discovery; and
it has been said, not without reason, that  ``he tumbled over it by accident''.
     Anyway, the discrepancy between Hansteen's detailed account and {\O}rsted's omissive one - ``effect so confused'' - of the famous lesson is irreconcilable. Kipnis even suggests that Hansteen saw nothing and that he simply imagined how a laboratory lesson on the discovery of {\O}rsted could be carried out \cite[p. 9]{kipnis}.
     \par
     It seems quite clear that, in these circumstances, it is appropriate to resort to epistemological criteria to find out what {\O}rsted has discovered. If we agree that the {\em discovery of a phenomenon} presupposes:
\begin{enumerate}
  \item the conscious search for the phenomenon; or its recognition if unexpected \label{consapevole}
  \item  the actual observation of the phenomenon \label{reale}
  \item the subsequent confirmation of the existence of the phenomenon \label{successiva}
\end{enumerate}
 there can be no doubt that -- at the end --  {\O}rsted actually {\em looked for and observed the deviation of a magnetic needle by a wire connected to a voltaic cell}. That the path that led {\O}rsted to the discovery was long, uncertain and rhapsodic, is reasonably established; that {\O}rsted set up an experiment by placing a magnetic needle near a conductor wire connected to a voltaic pile is beyond any reasonable doubt; that this experiment gave an uncertain outcome during the famous lecture is admitted by {\O}rsted himself. When {\O}rsted repeated and extended his lesson's experiment,
 a set of favorable factors not under the complete control of the experimenter played a fundamental role: sufficient current intensity due in turn to an appropriate combination of the electromotive force of the battery and the resistance of the wire, associated to an appropriate sensibility of the magnetic needle. In this sense, Kipnis  speaks of the role of casuality, meanwhile distinguishing this role from the concept of accidental discovery. However, this type of casuality occurs  in many experimental situations.
\subsection{{\O}rsted's report of the discovery}
{\O}rsted writes:
\begin{quote}\small
The first experiments respecting the subject which I mean at present
to explain, were made by me last winter, while lecturing on electricity,
galvanism, and magnetism, in the University. It seemed
demonstrated by these experiments that the magnetic needle was
moved from its position by the galvanic apparatus, but that the
galvanic circle must be complete, and not open, which last method
was tried in vain some years ago by very celebrated philosophers.
But as these experiments were made with a feeble apparatus, and
were not, therefore, sufficiently conclusive, considering the importance
of the subject, I associated myself with my friend Esmarck
to repeat and extend them by means of a very powerful galvanic
battery, provided by us in common
 \cite[p. 273, engl. transl.]{oersted}.
   \end{quote}
According to {\O}rsted, these experiments have been carried out in the following way. First, the magnetic needle was left to orient in the Earth's magnetic field. Then a straight metal wire connected to a voltaic pile was brought near the needle and parallel to it: the needle's deviation towards the  West or the East depended on the position of the wire, above or under the needle and on the direction of the current flow. The rotation of the wire's direction in a plane parallel to the South - North direction influenced the amount of the needle deviation; this depended also on the power of the pile and on the distance between the wire and the needle. From these observations, {\O}rsted concluded that it is no matter of attraction or repulsion between the wire and the needle, but that there must be some rotational effect involved in the interaction.
The action of the wire on the needle was not shielded by a series of interposed materials like glass, metal, wood, water and others. Also, the effect did not depend on the material of the wire. {\O}rsted, attributed the effect to an `electric conflict' which takes place in the conductor and in the surrounding space. {\O}rsted conceived what we now call an electric current as an {\em electric conflict}, namely  as a propagation of decomposition and recomposition of the two electric fluids; this electrical conflict somehow diffuses in the surroundings. Of course, this `electrical conflict' is little more than applying a name to a thing.
\subsection{Amp\`ere: a theory uniquely deduced from experiment?}\label{marie}
{\O}rsted's experiment could be easily reproduced: it required only a pile, a metal wire and a compass, at disposal in every laboratory.  Particularly reactive  were the French scientists: among them, Jean - Baptiste Biot (1774 - 1862), F\'elix Savart (1791 - 1841) and Andr\'e Marie Amp\`ere (1775 - 1836).  Biot and Savart (October, 1820) refined {\O}rsted's experiment giving it a quantitative expression:
\begin{quote}\small
    Using these methods, MM. Biot and Savart were led to the following result
which rigorously expresses the action experienced by a molecule of austral or boreal magnetism
placed at any distance from a very fine straight cylindrical wire, {\em made magnetic}
by the voltaic current. Draw from the pole a straight line perpendicular to
the axis of the wire: the force exerted on  the molecule is perpendicular to this line and to the axis of
wire. Its intensity is proportional to the reciprocal of the distance \cite[p. 223;  italics mine]{biot}.
\end{quote}
The expression `made magnetic' suggests the idea that the current in some way endows the wire with magnetic properties, responsible for the action on the magnetic needle.
Soon after, Amp\`ere showed that two parallel current carrying wires attract or repel each other according to the direction (parallel or antiparallel) of the currents. This  experiment   could have been  conceived only outside the common guess: if a current acts on a magnet, perhaps a magnet acts on a current. Indeed, {\O}rsted's  discovery suggested to Amp\`ere that the Earth magnetism -- which orients a magnetic needle -- could be due to currents flowing under the Earth's crust  along the Earth's parallels. From here, the step towards the hypothesis that the magnetic properties of magnets are due to native internal currents (molecular currents); these currents are there also in magnetizable material, the magnetization consisting in the ordering of otherwise disordered configurations of molecular currents. This bold, creative hypothesis allowed to unify under the same principle apparently different phenomena: the interactions between magnets and currents reduce to the interactions between currents. This hypothesis shaped Amp\`ere's later experiments and their interpretation.
\par
  All these phenomena were about forces exerted by something on something else. For  Amp\`ere, this mechanical phenomenology should be described  by the same mathematical tools used by   Newton for the  gravitational attraction and by Coulomb for the interaction  between electrical charges or between the poles of magnets.
  \par
  The title of Amp\`ere's most famous work, {\em Mathematical theory of electro - dynamic   phenomena uniquely deduced from experiment} appears as a manifesto of how to do physics on Newton's track \cite{amp2}. However, this title - manifesto is an unreliable epistemological account  of Amp\`ere's work. In the incipit,  Amp\`ere writes:
\begin{quote}\small
First observe the facts, vary the circumstances as much as
is possible, accompany this first work with precise measurements
to deduce general laws based solely on experiment,
and deduce from these laws, regardless of any assumption
on the nature of the forces that produce the phenomena, the mathematical value of these forces, that is to say the formula which represents them,
such is the course which Newton followed. This course was, in general, adopted
in France by the scientists to whom physics owes the immense
progress made in recent times, and  it is the one who
served as a guide in all my research on electrodynamic phenomena.
    I only consulted experience to establish the
laws of these   phenomena, and I deduced the only formula which can
represent the forces to which they are due; I did not do a
research into the very cause that can be assigned to these forces, well
convinced that any research of this kind should be preceded by
a purely experimental knowledge of laws, and  by a determination,
deduced only from these laws, of the value of the elementary forces
 whose direction is necessarily that of the straight line connecting
 the material points between which they are exerted \cite[p. 2 ]{amp2}.
\end{quote}
It is true that Amp\`ere wrote the formula of the force exerted by one circuit on another on the basis of experimental data taken with precise measurements carried out by varying the circumstances as much as possible. But he did so, within the basic hypothesis  that these forces are directed along the straight line joining two elementary portions of the interacting circuits and that these forces obey Newton's third principle, namely, that they are equal and opposite.   Moreover, Amp\`ere thought over what a current might be in terms of electric fluids. However, his idea of current could not be expressed in  mathematical form, susceptible of experimental test.  On the other hand, the fundamental role of theoretical arguments in framing the experimental data is acknowledged by Amp\`ere when, in dealing with his hypothesis of molecular currents in magnets, he affirms that its reliability is not due to a specific experimental corroboration but to its power of unifying different phenomena:
\begin{quote}\small
    The proofs on which I base it, result primarily from the fact that they reduce to a single principle three sorts of actions which all
phenomena prove to depend on a common cause, and which cannot be reduced [to a common cause] in
a different way \cite[p. 83 - 84]{amp2}.
\end{quote}
Here, the three `sorts of actions' are, of course,  that of a magnet on another, that of a current on a magnet (and viceversa) and that of a current on another one.
\par
The fundamental formula written by Amp\`ere consists in the expression of the force due to the interaction between two infinitesimal elements of current carrying circuits. Its laborious formulation by Amp\`ere  has been reconstructed, among others,  by Whittaker  \cite[pp. 87 - 92]{edmund}, Darrigol \cite[pp. 6 - 30]{darrigol}, Tricker \cite[pp. 46 - 55]{tricker},  Graneau  \cite[pp. 459 - 465]{graneau} and, extensively by Assis and Chaib \cite[pp. 27 - 140]{assis}. In modern notation, it can be written as \cite[pp. 27 - 29]{assis}, \cite[p. 357]{greco}:
\begin{equation}\label{formula!}
d^2\vec F_{12}= - \frac{\mu_0}{4\pi}I_1I_2 \frac{\vec r_{12}}{r_{12}^3}\left[ 2(\vec{dl_1}\cdot\vec{dl_2})- \frac{3}{r_{12}^2}(\vec {dl_1}\cdot \vec r_{12}) (\vec {dl_2} \cdot {\vec r_{12}})\right]
\end{equation}
where $d^2\vec F_{12}$ is the force exerted on the element $\vec {dl_2}$ by the  element $\vec {dl_1}$ and $\vec r_{12}$ is the position vector  of the element $dl_2$ relative to the element $dl_1$.
This formula is usually compared with the one, usually referred to as the Biot - Savart law:
\begin{equation}\label{biotsavart}
  d^2\vec F_{12}=   \frac{\mu_0}{4\pi}I_1I_2 \frac{1}{r_{12}^3}\vec {dl_2}\times (\vec {dl_1}\times \vec r_{12})
\end{equation}
These two formulas are different in two basic aspects: in Amp\`ere's formula, the force  is directed along the straight line connecting the two current elements, while in Biot - Savart's  the force is perpendicular to the current element;  Amp\`ere's formula is symmetrical with respect the two current elements,  Biot - Savart's  is not. However, when the integral forms of the two formulae are considered, i.e. when the force exerted by one circuit on the other (and viceversa)  is considered, the two formulae yield the same result. Both formulae yield the same result also for the force exerted by a complete circuit on an element of another circuit \cite[pp. 55 - 61]{tricker},  \cite[pp. 361 - 362]{greco}, \cite{ternan}. Of course, in this last case, Amp\`ere's formula yields a force that is perpendicular to the current element, as Biot - Savart's does.
\par
 This is one of the cases in which two different theories yield the same predictions testable by experiment. In such cases, the choice between the two theories is made on the basis of epistemological criteria. Amp\`ere's formula has been constructed on the basis of experiments within a Newtonian theoretical framework and does not belong to a general theory of electromagnetic phenomena. Instead, Biot - Savart's formula can be deduced within an axiomatic development based on Maxwell equations and the expression of the Lorentz force \label{biot} (\ref{pointcharge}). Hence, nowadays, Amp\`ere's formula has only a historical interest unless one is willing to look for a theory  alternative to MLE electromagnetism \cite{graneaubook}.
\par
 A very different fate has been that of Amp\`ere's hypothesis of molecular currents. In spite of its qualitative feature, Amp\`ere's hypothesis has been quantitatively embedded  in standard Electromagnetism,  by translating the magnetic properties of magnetic materials into  volume and superficial  current densities, through the equations:
 \begin{eqnarray}
   \vec J_m &=& \nabla \times \vec M\\
   \vec J_s &=& \vec M \times \hat n
 \end{eqnarray}
 where $\vec M$ is the magnetic moment per unit volume. It is an ironic twist of fate that -- in spite of Amp\`ere's Newtonian epistemology -- the part of his theoretical work which have survived is the one based on his bold, creative hypothesis and not the other, masterfully cast in mathematical form.
 \section{Electromagnetic induction: an unseen guest in the laboratories (1820 - 1830)}\label{unseen}
{\O}rsted's discovery triggered a series of experiments around Europe. In some of them, electromagnetic induction was there, but it was not noticed or  was misinterpreted. Amp\`ere was on the forefront.
In July 1821 he carried out an experiment to test whether a current carrying circuit can ``produce by influence'' a current in another circuit. Amp\`ere never used, in those years, the verb ``to induce'' or the term ``induction''. However, for sake of brevity, we will use them as synonyms of those used by Amp\`ere.
If the problem is posed in these terms, its formulation is ambiguous; and the answer of the experiment can only be equally ambiguous. In fact, for Amp\`ere, there are two types of current which we can denote as \textsl{voltaic current}, i.e. produced by a voltaic pile and as \textsf{molecular current}, responsible for the behavior of magnets and magnetizable materials.
Amp\`ere suspended, with a ``very fine metal'' wire, a copper ring inside a fixed coil connected to a voltaic pile and placed a magnet (of what type?) near the ring (how?).
If \textsf{molecular currents} were induced in the copper ring, this would have acquired magnetic properties, detectable  by a magnet. However, the magnet would have detected also the induction of a \textsl{voltaic current}. The experiment was intrinsically ambiguous, because to the experimenter's question `there is something induced' the answer could be `yes, \textsf{molecular currents}' or  `yes, \textsl{voltaic current}'. This ambiguity was never explicitly dissolved by Amp\`ere before the discovery by Faraday of electromagnetic induction.
Furthermore, we must take into account that Amp\`ere's hypothesis of \textsf{molecular currents} did not imply the possibility of creating them in non magnetizable materials like copper. In other words, Amp\`ere's expectation was for a negative result of the experiment.
\par
 In a letter sent to  Albert van Beck in 1822, Amp\`ere, referring to this experiment, writes:
 \begin{quote}\small
 Anyway, I have carried out in July 1821 an experiment which  proves    only indirectly that   \textsf{electric currents} [molecular] in the magnet take place around
   each molecule.
Instead, this experiment proves directly that the proximity of an \textsl{electric current} [voltaic] does not excite at all any [molecular or voltaic] by influence, in a copper metallic circuit, even under the best conditions for such an influence \cite[p. 447 - 448, fonts and labels in square bracktes mine]{beck}.
 \end{quote}
 This statement is ambiguous, because of the use of the same term `electric current' for designating both \textsf{molecular currents}
and \textsl{voltaic currents}.   In the passage quoted above,  the word `any' can be interpreted as referring to  \textsf{molecular currents}. This interpretation offers a consistent reconstruction of Amp\`ere's  interpretation of the experiment and of his reaction to the positive result of a second experiment performed one year later in Geneva, where a stronger horse magnet was at disposal. Of course, also the other choice is  legitimate; however, it  leads to the conclusion that Amp\`ere completely overlooked  the importance of the positive result of the second experiment.
\par
  The only published report of  this second experiment was that of Auguste de la Rive (1801 - 1873), Amp\`ere's young collaborator in Geneva:
 \begin{quote}\small
By presenting a very strong horseshoe magnet to one side of this ring, we saw it sometimes advance
between the two branches of the magnet, sometimes to be repelled, according to the direction of the current
in the surrounding conductors. This important experiment
therefore shows that bodies which are not susceptible,
through the influence of electric currents,
to acquire permanent magnetization, such as
 iron and steel, can at least acquire a
sort of temporary magnetization while they are under this influence \cite[p. 47 - 48]{rive}.
 \end{quote}
 de la  Rive's report, speaking of `temporary magnetization' of the ring, supports the interpretation according to which the currents induced in the ring are molecular currents. In other words, the copper ring acquires transient properties typical of a magnet, owning to the induction of \textsf{molecular currents}.
In a report's draft, written just after his return to Paris but never published during his life, Amp\`ere writes:
 \begin{quote}\small
    The closed circuit placed under the influence of the redoubled electric current, but without any communication with it, was attracted and repelled alternately by the magnet, and this experience would therefore leave no doubt about the production of electric currents by influence, if we could not suspect the presence of a little iron in the copper from which the circuit was formed.  However, there was no action between this circuit and the magnet before the electric current passed through the spiral  which  surrounded it; this is why I regard this experiment as sufficient to prove this production. Nevertheless,  to prevent any objection, I plan to repeat it immediately, with a circuit made of highly purified  non - magnetic metal. The fact  that electric currents can be produced by influence is very interesting in itself, and  is besides independent of the general theory of electrodynamic action (Quoted, in original French language by \cite[p. 65]{uffa}).
 \end{quote}
 Amp\`ere speaks of induced `electric currents' without specification; however, his subsequent comment about the possible presence of iron in the copper,  suggests that he his thinking of \textsf{molecular currents}, thus confirming de la Rive's interpretation in terms of magnetization of the ring.  The last comment on the independence of the experimental result from his theory suggests that  Amp\`ere's main concern was to preserve his electrodynamic theory, molecular currents included. If  Amp\`ere would have found an induced \textsl{voltaic current} instead of induced \textsf{molecular currents}, he very likely would have realized the absolute novelty of the discovery and he would have thoroughly investigated the new phenomenon. He did not, thus leaving the state of the experimental knowledge limited to the superficial  observations reported by de la Rive and himself.
 In particular, both reports do not give any indication of a transient effect due to the switching on and off of the current in the inducing circuit.
  Since this effect was the only one present, it is in principle possible that the copper ring behaved as an overdamped ballistic galvanometer, keeping its deviation for a time interval much greater than that necessary for establishing the stationary current in  the inducing circuit. This is the conclusion reached by a recent reproduction of the experiment \cite{mendoza}.
    Though reproductions of historical experiments can  illustrate the difficulties encountered by the original authors, it is questionable wether they  can significantly  contribute  to the historical reconstructions of  past events, when the experimenters' reports are inaccurate as in this case.  In fact, de la Rive and Amp\`ere do not speak of an almost stationary deviation of the galvanometer, nor of a transient one. They only reported of a deviation.
    This remind us of the epistemological criteria used above for {\O}rsted's discovery (page \pageref{consapevole}). Amper\`e and de la Rive observed a deviation of the ring. But the deviation should have been transient. Since this transient effect has not been reported, it is  clear that the experimenters did not ``actually observed the phenomenon'', thus leaving no doubt that they completely missed the discovery of electromagnetic induction. For detailed discussions of Amper\`e's `induction' experiment see, for instance,
    \cite{uffa, ross, williams, devons}.
 Therefore, it is not surprising that,
after the discovery of electromagnetic induction by Faraday in 1831,   Amper\`e's attempt to vindicate the discovery   failed.  In a paper published in 1831, immediately after learning of Faraday's discovery, Amper\`e specified that, in the Geneva's experiment:
\begin{quote}\small
We presented
to this ring a strong horseshoe magnet,
so that one of the poles was inside and the other
outside the ring. As soon as we connected
 the two ends of the conducting wire to the poles of the battery, the
ring was attracted or repelled by the magnet, depending on the
pole that was inside the ring; which demonstrated
the existence of the electric current
produced by the influence of the current of the conducting wire \cite[p. 405]{amp1831}.
\end{quote}
This important detail was lacking in de la Rive's and Amper\`e's reports. Furthermore, a statement about what happened to the ring after the establishment of the induced current in it, was still missing.
It is possible that -- in the Geneva's experiment -- they did see a transient deviation of the ring; but to have dismissed this observation as insignificant was a gross mistake. If we take  Amper\`e's belated reconstruction of  Geneva's experiment as trustworthy, we are again led to the conclusion that,  in the first Twenties,  Amper\`e's main concern was that of preserving his theory of electric currents, molecular currents included. For an in - depth analysis of Amper\`e's 1831 paper and  the ensuing correspondence with de la Rive and Faraday, see \cite[pp. 209 - 213]{ross}.
\par
In September, 1821,  Michael Faraday showed that the action of a magnet on a current carrying wire can produce a continuous rotatory motion of the wire around the magnet's axis \cite{faradayQ}. Here, the electromagnetic induction is a secondary effect, masked by the primary one. We now know that when the wire is set in motion,  an induced current flows in the wire in the opposite direction of the primary current: the induced current tends to slow down the motion of the wire. In a stationary condition, the electric energy furnished by the pile is distributed between the wire's kinetic energy and the losses due to the Joule effect.
\par
On November 22, 1824 Fran\c{c}ois Arago reported to the French Academy that a rotating copper disc produces a deflection of a magnetic needle suspended above it in the same direction of disc's rotation. If the rotation speed is high enough, the needle follows the disc in its rotation. A detailed account of this experiment, appeared only two years later, when Arago entered a controversy  with Liberato Baccelli (1772 - 1835) and Leopoldo Nobili (1784 - 1835) \cite{arago1826}. At that time, Arago's experiment had already been carefully repeated and extended by Charles Babbage (1791 - 1871) and Frederick William Herschel (1738 - 1822) \cite{babbage}: they demonstrated also the reverse phenomenon by rotating a horseshoe magnet around its axis below a suspended copper disc. This astonishing phenomenon, which we now know as due to the eddy currents induced in the disc, could not find an explanation in those times.
\section{Electromagnetic induction: the discovery}\label{discovery}
On November 24 of the year 1831, Michael Faraday presented to the Royal Society of London a research entitled `Induction of electric currents'. In May, 1832 Faraday's paper on electromagnetic induction was published in the {\em Philosophical Transactions} and was to open the first volume of his monumental {\em Experimental Researches in Electricity} \cite{faraday1} \footnote{Faraday's paper was among the firsts to be submitted to referees before publication in the {\em Transactions} \cite{referees}. }.
The experiments carried out by Faraday constitute the basic experimental evidence of currents induced in a closed conducting loop by switching on and off the current in another nearby circuit (volta - electric - induction) \cite[pp. 1 - 7]{faraday1}, or by moving a loop  towards or away from a magnet (magneto - electric - induction) \cite[pp. 7 - 16]{faraday1}. These experiments are usually presented in  textbook or classrooms as the starting point for talking about  electromagnetic induction.
Joseph Henry had obtained some of these  results one year before, but  published them one year later  \cite{henry}.
\par
Faraday's researches can be considered as an archetype of experimental enquire. A significant experiment, namely an
experiment capable of producing new knowledge, stems from the acquired knowledge
 and is designed to `interrogate the Nature' about an unresolved issue.
The idea of the experiment can not be logically deduced  from the
acquired knowledge, but requires a creative effort
consisting of a hypothesis on `how things should go
in the world'. Then, the experimenter must  evaluate
if the result  supports or falsifies the starting hypothesis. In the latter case, the hypothesis must be changed
and, consequently, a new experiment must be devised. This procedure is systematically and
masterfully applied in the {\em Experimental Researches}. With the words of
Maxwell: ``The method which Faraday employed in his researches consisted in a constant appeal to experiment as a mean of testing the truth of his ideas, and a constant cultivation of ideas under the direct influence of experiment''  \cite[p. 162, \S 528]{treatise2}.
\par
Faraday did not receive a formal school education: in particular,
Faraday never learned to use the calculation tools offered by mathematics.
Nonetheless, while not using any formula, Faraday formulated quantitative laws and always sought coherent descriptions of phenomena that can be considered, to all effects,
 as  theories.
 This is authoritatively acknowledged by
 Maxwell, who  writes:
\begin{quote} \small
As I proceeded with the study of Faraday, I perceived
that his method of conceiving the phenomena
was also a mathematical one, though not exhibited
in the conventional form of mathematical symbols. I
also found that these methods were capable of being
expressed in the ordinary mathematical forms, and
thus compared with those of the professed mathematicians.
\cite[p. X]{treatise1}.
 \end{quote}
\subsection{Faraday's theory of electromagnetic induction}\label{fartheory}
Faraday sought to develop a theory of electromagnetic induction over more than twenty years. While his basic experiments date back to the first Thirties, his mature reflections have been developed in the Fifties.  We shall try to present Faraday's theoretical effort  by  taking into account comprehensively his reflections, independently from their  chronological order.  Furthermore, on the wake of Maxwell, we shall try to find out if and how Faraday's ideas can be cast in mathematical form.
\par
Faraday initially thought that the electromagnetic induction  was due to a particular state of matter named `electrotonic state' \cite[\S 60, p. 16]{faraday1}, but  he soon realized that this hypothesis was too generic for yielding any experimental prediction \footnote{In fact,  in the same communication, Faraday writes ``Thus the reasons which induced me to suppose a particular
state in the wire (\S 60) have disappeared\dots \cite[\S 242, p. 69]{faraday1}''. However, Faraday did not completely abandon the concept of electrotonic state. This concept reappears from time to time in his writings.}.  Therefore, he formulated the idea of magnetic curves or {\em lines of magnetic force} and found that this concept, conveniently used, could describe the observed phenomena.
The lines of magnetic force are [June, 1852]:
\begin{quote}\small\label{lines}
    \dots
those lines which are indicated in a general manner by the disposition of iron filings
or small magnetic needles, around or between magnets;
and I have shown, I hope satisfactorily, how these lines may be taken as exact representants of the magnetic power,
both as to disposition and amount;
also how they may be recognized by a moving wire in a manner altogether different in
principle from the indications given by
a magnetic needle, and in numerous cases
with great and peculiar advantage \cite[\S\, 3243, p. 407]{faraday3}.
\end{quote}\small
The concept of lines of magnetic force is  used for describing `the magnetic power', i.e. -- in our language -- the intensity (amount) and direction (disposition) of the magnetic field. Then, it can be translated into the language of fields by carefully substitute to them our $\vec B$.
\par
\begin{figure}[htb]
\centering{
 \includegraphics[width=3.5cm]{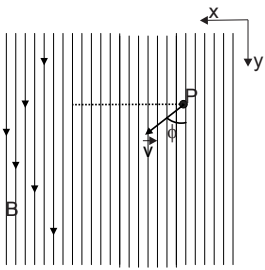}
 }
\caption{\label{farbax}  Point $P$ represents  a rigid conducting wire placed perpendicularly to the lines of magnetic force which describe a constant and uniform magnetic field.  If the velocity vector is parallel to the lines of magnetic force, the wire intersects no lines; if the velocity vector is perpendicular to the lines, the number of lines intersected is maximum. In general,  the number of lines intersected is $\propto v \sin \theta$.}
 \end{figure}

 In considering the motion of a wire in a constant and uniform magnetic field, Faraday writes [October, 1851]:
 \begin{quote}\small
 It is also evident, by the results of the rotation of the wire
and magnet (\S\, 3097 - \S\, 3106), that when a wire is moving amongst
equal lines (or in a field of equal magnetic force), and with an
uniform motion, then the current of electricity produced is proportionate
to the time; and also to the velocity of motion.
They also prove, generally, that the quantity of electricity
thrown into a current is directly as the amount of curves
intersected \cite[\S\,3114 - 3115, p. 346]{faraday3}.
 \end{quote}
Notice that here, Faraday uses the term `magnetic field' in the same meaning of Maxwell: the region of space where the magnetic forces operate. Let us try to illustrate this Faraday's passage with  figure  \ref{farbax}, where a wire, perpendicular to the lines of magnetic force, moves in an uniform and constant magnetic field.
According to Faraday, the electric charge injected in the induced current is proportional to the duration of the motion and to the velocity of the wire. Furthermore,  this same quantity is proportional to the number of lines of magnetic force $\Delta N$ intersected.
 We can translate Faraday's description in a formula by writing:
 \begin{equation}\label{farvb}
    \Delta q\propto v_\perp \Delta t\Delta N
 \end{equation}
 or in terms of the induced current:
 \begin{equation}\label{farivb}
    i=\frac{\Delta q}{\Delta t}\propto vB\sin\theta
 \end{equation}
 where $B$ is our magnetic field.
 \par\noindent
The translation of Faraday's experimental results into the language of  fields contains -- with hindsight -- some features of the magnetic component of  Lorentz force. On the other hand, it stresses the limits of Faraday's inability to use mathematical tools.
\par
The relative motion of conductors and lines of magnetic force  has been considered by Faraday also in a different and conceptually intriguing experimental situation [October, 1851]:
\begin{quote}\small
    When {\em lines of force} are spoken of as crossing a conducting
circuit (\S 3087), it must be considered as effected by the
{\em translation} of a magnet. No mere rotation of a bar magnet on
its axis, produces any induction effect on circuits exterior to it;
for then, the conditions above described (\S\, 3088) are not fulfilled.
The system of power about the magnet must not be considered as
necessarily revolving with the magnet, any more than the rays of
light which emanate from the sun are supposed to revolve with the
sun. The magnet may even, in certain cases (\S\, 3097), be considered
as revolving amongst its own forces and producing a full
electric effect, sensible at the galvanometer \cite[\S 3090, pp. 336 - 337, original italics]{faraday3}.
\end{quote}
Here, Faraday speaks of lines of force that cross the  conducting circuit, thus introducing the idea that the lines of magnetic force may move. However, this movement of the lines occurs only when the magnet is translated, not when it is rotated. In the language of fields, this simply means that, considered an arbitrary point outside the magnet, the rotation of the magnet does not change the value of the magnetic field, while its translation does.
\par
The idea of lines of magnetic forces that move contains implicitly the problem of the `velocity of the lines of magnetic force'.  Faraday speaks of it  with these words [October, 1850]:
\begin{quote}\small
    In space, I conceive that the magnetic lines of force, {\em not
being dependent on or associated with matter} (\S 2787, \S 2917),
{\em would have their changes transmitted} with the velocity of light,
or even with that higher velocity or instantaneity  which we suppose
to belong to the lines of gravitating force, and if so, then a
magnetic disturbance at one place would be felt instantaneously
over the whole globe \cite[\S 2958, p. 264; italics mine]{faraday3}
\end{quote}
In the above two quotations, we find two basic concepts: the {\em independency} of the lines of magnetic force from matter; and, consequently, that  their {\em changes} are transmitted with the velocity of light or instantaneously.
This  property of the lines of magnetic force allows Faraday to describe, at least qualitatively,  all the  studied phenomena.
In fact [January, 1832]:
\begin{quote}\small
    To prove the point with an ordinary magnet, a copper
disc was cemented upon the end of a cylinder magnet, with
paper intervening; the magnet and disc were rotated together,
and collectors (attached to the galvanometer) brought in contact
with the circumference and the central part of the copper plate.
The galvanometer needle moved as in former cases, and the
direction of motion was the same as that which would have resulted,
if the copper only had revolved, and the magnet been
fixed. Neither was there any apparent difference in the quantity
of deflection. Hence, rotating the magnet causes no difference
in the results; for a rotatory and a stationary magnet
produce the same effect upon the moving copper \cite[\S\,218, p. 63]{faraday1}.
\end{quote}
\label{non}According to Faraday, this  experiment  shows that the lines of magnetic force do not rotate with the magnet. When the disc and the magnet rotate together, there is no relative motion between them. Nonetheless, since there is induced current, we must conclude that there is relative motion between the disc and the lines of magnetic force, i.e. that the lines of magnetic force do not rotate with the magnet.
Likely for this reason, Faraday did not perform the experiment in which only the magnet rotates. This Faraday's explanation is translated into the language of fields by the treatment developed in section \ref{faruni}.
\par
Two paragraphs ahead, Faraday's writes (January, 1832):
 \begin{quote}\small
    That the metal of the magnet itself might be substituted
for the moving cylinder, disc, or wire, seemed an inevitable
consequence, and yet one which would exhibit the effects of
magneto  - electric induction in a striking form. A cylinder magnet
had therefore a little hole made in the centre of each end
to receive a drop of mercury, and was then floated pole upwards in the same metal contained in a narrow jar. One wire
from the galvanometer dipped into the mercury of the jar,
and the other into the drop contained in the hole at the upper
extremity of the axis. The magnet was then revolved by a
piece of string passed round it, and the galvanometer - needle
immediately indicated a powerful current of electricity. On
reversing the order of rotation, the electrical current was reversed.  The direction of the electricity was the same as if the
copper cylinder  (\S 219) or a copper wire had revolved round the
fixed magnet in the same direction as that which the magnet
itself had followed. Thus a {\em singular independence} of the magnetism
and the bar in which it resides is rendered evident \cite[\S 220, p. 64, original italics]{faraday1}.
 \end{quote}
  This phenomenon, which Faraday considered -- correctly --  as an ``inevitable
consequence'' of the experiment with the disc, was later labeled as `unipolar induction' \label{unipo}(or ``homopolar induction'') and gave rise to a theoretical dispute whose fragments have reached till our days (sections \ref{unisec}, \ref{why}).
\par
According to Faraday, the relative motion between conductor and lines of magnetic force
is also the cause of the current induced in a circuit by a variable current
 circulating in another circuit [January, 1832]:
\begin{quote}
    In the first experiments (\S 10, \S 13), the inducing wire
and that under induction were arranged at a fixed distance
from each other, and then an electric current sent through the
former. In such cases the magnetic curves themselves must
be considered as moving (if I may use the expression) across
the wire under induction, from the moment at which they begin
to be developed until the magnetic force of the current is at its
utmost; expanding as it were from the wire outwards, and
consequently being in the same relation to the fixed wire under
induction as if {\em it } moved in the opposite direction across
them, or towards the wire carrying the current. Hence the first
current induced in such cases was in the contrary direction to
the principal current (\S 17, \S 235). On breaking the battery contact, the magnetic curves (which are mere expressions for arranged
magnetic forces) may be conceived as contracting upon
and returning towards the failing electrical current, and therefore move in the opposite direction across the wire, and cause
an opposite induced current to the first  \cite[\S238, p. 68; original italics]{faraday1}.\label{vellines}
\end{quote}
Notice how Faraday deals differently with  the concept of the motion of  lines of magnetic force in 1832 and in 1850. In 1832, the lines of magnetic force  ``must
be considered as moving, (if I may use the expression)''. About twenty years later, the lines of magnetic force are clearly conceived as independent of matter and their changes are transmitted with the velocity of light, or instantaneously.
A \label{translation} translation into the  language of fields of this Faraday's description is carried out in terms of the vector potential (section \ref{mle}). The variation of the current in the inducing circuit produces at each point of the induced circuit an electric field  $\vec E=-\partial \vec A/\partial t$,  where $\vec A$ is calculated from the known values of the current circulating in the inducing circuit. Notice that the use of the vector potential  satisfies Faraday's requirement that the physical interaction with every point of the induced circuit must be local: this condition is not satisfied by the use of the magnetic field through the ``flux rule'' (section \ref{regola}).
\par
Finally, an epistemological note.  Speaking of lines of magnetic force that move, Faraday stresses that he is dealing  {\em only} with
a theoretical entity, without any ontological commitment about their existence in the world.
However, in the Fifties, Faraday returned to this issue, again with a strong methodological warning
[March, 1852]:
 \begin{quote}\small
    Having applied the term {\em line of magnetic force} to an
abstract idea, which I believe represents accurately the nature,
condition, direction, and comparative amount of the magnetic
forces, without reference to any physical condition of the force,
I have now applied the term {\em physical line of force} to include the
further idea of their physical nature. The first set of lines I
{\em affirm} upon the evidence of strict experiment (\S 3071 \& c.). The
second set of lines I advocate, chiefly with a view of stating the
question of their existence; and though I should not have raised
the argument unless I had thought it both important, and likely
to be answered ultimately in the affirmative, I still hold the
opinion with some hesitation, with as much, indeed, as accompanies
any conclusion I endeavour to draw respecting points in
the very depths of science, as for instance, regarding one, two
or no electric fluids; or the real nature of a ray of light, or the
nature of attraction, even that of gravity itself, or the general
nature of matter \cite[\S\,3299, p. 437, original italics]{faraday3}.
\end{quote}
Faraday clearly stresses the basic difference between a theoretical term and its use in a theory from the statement about its real existence in the world. While the former is subjected to experimental test through the predictions of the theory, the latter may be only a more or less plausible statement about how things go into the world.
\section{James Clerk Maxwell}\label{maxwell}
Maxwell's electromagnetic theory was developed within
an image of the world  in which the ether
played a key role. Maxwell himself wrote the entry {\em
Ether} for the ninth edition of the {\em Encyclop{\ae}dia Britannica}
 in which the electromagnetic field was conceived as a
mechanical deformation of the ether \cite{etere}. Nonetheless, Maxwell's theory survived the emergence of the discrete nature of the electric charge, the disappearance of the ether,  the birth of  special relativity and  the resurgence
of the corpuscular description of light. The intimate reason lies in the fact that,
as
 Hertz  put it with a conscious
simplification,  ``Maxwell's theory is Maxwell's system of equations
\cite [p. 21] {ew}''.
 \par
  The birth of
special relativity created no problem for the simple fact that Maxwell's  is a relativistic theory.
Indeed, Maxwell's theory entered implicitly in the  axioms of special relativity  through Einstein's postulate
 according to which ``light in empty space always propagates with one
speed determined $ c $, independent of the state of motion of the bodies
issuers \cite[engl. trans. p. 100]{ein05lq}''.
 This postulate can be replaced by: Maxwell's equations in vacuum are true. Einstein dedicated to this issue a note in a subsequent paper: ``The principle of the constancy of the velocity of light used there [the paper on special relativity] is of
course contained in Maxwell's equations'' \cite[engl. trans., p. 172]{ein05inerzia}.  Indirectly, Maxwell's Electromagnetism was therefore the cause of the reduction of  Newtonian dynamics to an approximation of relativistic dynamics for low velocities. Furthermore, the predictions of Maxwell's Electromagnetism are still valid when the number of photons used is high enough. For instance, this is true for the interference experiments carried out with one photon at a time \cite[p. 564]{singolo}. In particular, the fact that the probability of finding a photon at a point on the detector is proportional to the electromagnetic intensity at the same point is due to the same mathematical structure of the electromagnetic and quantum description \cite[pp. 12 - 15 ]{luce} \cite[pp. 166 - 170]{erq}.
 Finally, Maxwell's equations for vacuum and without
sources, with the electric and
magnetic fields replaced by suitable operators, are at the
base  of Quantum Electrodynamics. Perhaps, in the
history of physics there is no other theory which has remained vital and creative through so many
radical transformations.
\subsection{Maxwell and the electromagnetic induction}\label{maxind}
In the introductory and descriptive part of the {\em Treatise} dedicated to
electromagnetic induction \cite[pp. 163 - 167]{treatise2}, Maxwell writes:
\begin{quote}\small
    The whole of these phenomena may be summed up in one
law. When the number of lines of magnetic induction  which pass
through the secondary circuit in the positive direction is altered,
an electromotive force acts round the circuit, which is measured
by the rate of decrease of the magnetic induction through the
circuit  \cite[\S 531, p. 167]{treatise2}.
\end{quote}
In  formula:
\begin{equation}\label{flusso}
    \mathcal E=-\frac{d\Phi(B)}{dt}
    \end{equation}
Maxwell does not write this formula, that, instead, can be derived  by combining two equations written some pages ahead; see below.
Equation (\ref{flusso}) is the  ``flux rule'', which is generally considered as the {\em law} of electromagnetic induction.
However, Maxwell wrote a {\em general law of electromagnetic induction}. In order to discuss this  issue, we need to take a few steps back.
\par
As we have seen, physicists learned  how to describe mathematically the interactions between current carrying wires  without knowing
what a current was. This ignorance did not imped Georg Simon Alfred Ohm (1789 - 1854) to establish his laws (in modern notation along with Ohm's):
 \begin{eqnarray}
   i &=&  \frac{\mathcal E}{r+R};\qquad X =\frac{a}{b + x}\label{seconda}\\
   i&=&\sigma \frac{A}{l}\Delta V; \qquad X=k \frac{\omega}{l}a \label{prima}
 \end{eqnarray}
 ``where $X$ is the strength of the magnetic action
of the conductor whose length is $x$, and $a$ and $b$
are constants depending on the exciting force
 and resistance of the other
parts of the circuit \cite[p. 151]{ohm26}.'' Notice that Ohm speaks of the magnetic action of  `the conductor' and not of the magnetic action of the `current' and that Ohm measured the current intensity through the deviation of a magnetic needle in a torsion balance. Ohm established his laws by using a Cu - Bi thermocouple as a current source, instead of the  voltaic pile used in previous measurements. This choice was forced by the instability of the currents produced by a voltaic pile. For discussions of these topics, see, for instance, \cite{pourprix, paoloohm}.
These achievements took a long time to be recognized, owing to the poor knowledge of the properties of the voltaic pile, in particular of the role played by its internal resistance $r$ ($b$ in Ohm's) and owing to the laborious process of clarification of the meaning of the {\em emf}  $\mathcal E$ ({\em exciting force} $a$ in Ohm's) and current intensity $i$ ($X$ in Ohm's). Ohm's main paper appeared in 1827 \cite{ohm27}, but the acceptance of his laws required many years, as documented also by the delay with which his paper was translated in English (1841) \cite{ohm41}.
\par
Maxwell acknowledges his ignorance about what a current is with these words:
\begin{quote}\small
    The electric current cannot be conceived except as a kinetic
phenomenon.
\par
[\dots]
\par
The effects of the current, such as electrolysis, and the transfer
of electrification from one body to another, are all progressive
actions which require time for their accomplishment, and are therefore of the nature of motions.
\par
{\em As to the velocity of the current, we have shewn that we know
nothing about it, it may be the tenth of an inch in an hour, or
a hundred thousand miles in a second. So far are we from
knowing its absolute value in any case, that we do not even know
whether what we call the positive direction is the actual direction
of the motion or the reverse}.
\par
{\em But all that we assume here is that the electric current involves
motion of some kind}. That which is the cause of electric currents
has been called Electromotive Force. This name has long been
used with great advantage, and has never led to any inconsistency
in the language of science. Electromotive force is always to be
understood to act on electricity only, not on the bodies in which
the electricity resides. It is never to be confounded with ordinary
mechanical force, which acts on bodies only, not on the electricity
in them. If we ever come to know the formal relation between
electricity and ordinary matter, we shall probably also know the
relation between electromotive force and ordinary force \cite[\S 569, pp. 196 - 197; italics mine]{treatise2}.
\end{quote}
The lacking knowledge was that  electricity is a property of  particles endowed with mass. Then, Maxwell could have written $F=qE=ma$, thus unveiling the `the formal relation between
electricity and ordinary matter'.
\par
The idea that the current is a kinetic phenomenon,  allowed Maxwell to treat electrical circuits with the Lagrangian formalism and to use a mechanical analogy. Considered a systems of two electrical circuits, he writes that the kinetic energy of the system due to its electrical properties -- i.e its {\em electrokinetic energy} --  is given by \cite[\S 578, p. 207]{treatise2}:
\begin{equation}\label{cincircuiti}
    T=\frac{1}{2}L_1{\dot{y_1}}^2 + \frac{1}{2}L_2{\dot{y_2}}^2 + M_{12}\dot{y_1}\dot{y_2}
\end{equation}
where $\dot{y_1}$ and $\dot{y_2}$ are the currents in the circuits. In $L_1, L_2$ and $M_{12}$ we recognize, as Maxwell does, the inductances of the two circuits and their mutual inductance \footnote{The  analogy between a coil's self - inductance  and the inertial mass of a particle is underlined  by Feynman in his {\em Lectures} \cite[pp. 17,11 - 17,12]{feyn2}.  Feynman considers an ideal inductance $L$ inserted in a circuit. The voltage difference between the two ends of the inductance is given by: $V(t)=L(dI/dt)$. This equation has the same form of $F=m(dv/dt)$. Hence, the analogy between the self - inductance and the inertial mass of a particle and that between the current and the velocity. Of course, the epistemological stand of the two analogies is very different. For Maxwell, the analogy is a creative one, because it leads to the general law of electromagnetic induction. Instead, for Feynman, the analogy is only an ex post consideration. [Thanks to Biagio Buonaura for having recalled my attention to Feynman's analogy.]}. Then, the {\em electrokinetic momentum} of, for instance, circuit $2$ is given by:
\begin{equation}\label{momele}
    p_2=\frac{dT}{d\dot{y_2}}= L_2\dot{y_2}+M_{12}\dot{y_1}
\end{equation}
The physical dimensions of $T$ are those of an energy;  the {\em electrokinetic momentum}, namely the analog of mechanical momentum,  has instead the dimensions  of an electric  potential multiplied by time.
The  role played by  $T$ appears clearly when a single, isolated (non interacting with other circuits), circuit is considered. In this case, and with our notations:
\begin{equation}\label{isolato}
    T= \frac{1}{2}Li^2;\qquad p=Li
\end{equation}
Then, $T$ is the magnetic energy of the magnetic field created by the current $i$ and $p=Li=\Phi$, where $\Phi$   is the magnetic flux through the circuit.
Since $\Phi =\oint \vec A\cdot \vec{dl}$, where $\vec A$ is the vector potential, we get for the {\em electrokinetic momentum}:
\begin{equation}\label{vecpot}
    p=\oint \vec A\cdot \vec{dl}
\end{equation}
Let us now consider a conducting filiform loop that, at the instant $t=0$ is connected to the poles of a battery. As  we now explain to our students, the current which flows in the loop does not attain at once its stationary value $\mathcal E/R$, owing to the {\em emf} self - induced in the loop, {\em emf} that contrasts the current increase. The energy supplied by the battery during this process is:
\begin{equation}\label{induttanza}
    E=\int_{0}^{\infty}{Ri^2dt}+\frac{1}{2}LI^2
\end{equation}
The first term is the energy dissipated in the circuit as heat, the second is the energy $T$ of Maxwell's equation (\ref{isolato}) in steady conditions.
Instead of considering the energy balance, Maxwell writes \cite[\S 579, p. 208]{treatise2}:
\begin{equation}\label{indumax}
    \mathcal E= R\dot{y}+ \frac{dp}{dt};\qquad  \mathcal E= Ri+L\frac{di}{dt}
\end{equation}
where we have added our translation of Maxwell's equation.
Maxwell's comment:
\begin{quote}\small
    The impressed electromotive force $\mathcal E$ is therefore the sum of two
parts. The first, $R\dot{y}$, is required to maintain the current $\dot{y}$ against
the resistance $R$. The second part is required to increase the electromagnetic momentum $p$.
This is the electromotive force which
must be supplied from sources independent of magneto - electric
induction. The electromotive - force arising from magneto - electric
induction alone is evidently $-{dp}/{dt}$, or, {\em the rate of decrease of the
electrokinetic momentum of the circuit} [original italics].
    \end{quote}
Therefore,  the {\em emf} induced in an isolated circuit is given by:
\begin{equation}\label{emfdef}
    \mathcal E= -\frac{dp}{dt}=-\frac{d}{dt}\oint_l \vec A\cdot \vec{dl}
\end{equation}
where $\mathcal E$ has the dimensions of an electric potential, as it should.
\par
Naturally, equation (\ref{emfdef}) is valid also in the case of two or more circuits, as it can be easily verified.
In a section entitled {\em Exploration of the field by means of the secondary circuit} \cite[p. 212]{treatise2}, Maxwell treats in detail the case of two circuits in several conditions.
Maxwell begins by recalling that, on the basis of equation (\ref{momele}) the electrokinetic momentum of the secondary circuit consists of two parts; the interaction part is given by:
\begin{equation}\label{momesec}
    p_2= Mi_1
\end{equation}
Under the assumptions that {\em the primary circuit {\em (circuit $1$)} is fixed and its current $i_1$ is constant},  the electrokinetic momentum of the secondary circuit depends only -- through the mutual inductance $M$ --  on its form and position. Under the above conditions \cite[\S 592, p. 216]{treatise2}:
\begin{equation}\label{flusso2}
 \oint_{l_2}    \vec A\cdot \vec{dl_2} = \oint_{S_2} \vec B \cdot \hat n\, dS_2
\end{equation}
where $\vec B$ is the magnetic field.
Now, if the secondary circuit is at rest, by combining equations (\ref{emfdef}) and (\ref{flusso2}), we get the ``flux rule'' (\ref{flusso}). The fact that Maxwell does not make this step, confirms that, in his mind, the basic law of electromagnetic induction is given by $\mathcal E=-dp/dt$.
\par
Indeed, this equation is the starting point for obtaining  the {\em General Equations of the Electromotive Force} \cite[\S 598, p. 220]{treatise2}.
Maxwell considers the secondary circuit in motion with a velocity vector for each circuit element: this means that the circuit may also change form. The {\em emf} induced in the secondary circuit is given by:
\begin{equation}\label{partenza}
    \mathcal E= -\frac{dp}{dt}=-\frac{d}{dt}\oint_{l_2}\vec A(\vec r, t)\cdot \vec{d{l_2}}
\end{equation}
The details of the calculation (omitted by Maxwell) can be found in  \cite{max_598}. It turns out that (in modern notation):
\begin{equation}\label{leggegenmax}
      \mathcal E= -\frac{dp}{dt}=\oint_{l_2} \left[(\vec v \times \vec B) -\frac{\partial \vec A}{\partial t} - \nabla\varphi \right]\cdot \vec {dl_2} =
      \oint_{l_2} \vec E_v\cdot \vec {dl_2}
      \end{equation}
      where we have added the suffix $v$ to the vector $\vec E$, not present in Maxwell's text.
Maxwell's comment:
\begin{quote}\small
    The vector $\vec E_v$ is the electromotive force [electric field] at the moving element
$dl_2$.
\par
[\dots]
\par
The electromotive force [electric field] at a point has already been defined in
\S 68. It is also called the resultant electrical force, being the
force which would be experienced by a unit of positive electricity
placed at that point. We have now obtained the most general
value of this quantity in the case of a body moving in a magnetic
field due to a variable electric system.
If the body is a conductor, the electromotive force [electric field] will produce a
current; if it is a dielectric, the electromotive force [electric field] will produce
only electric displacement. The electromotive force [electric field] at a point, or on a particle, must be
carefully distinguished from the electromotive force along an arc
of a curve, the latter quantity being the line - integral of the former.
See \S  69. \cite[\S 598, p. 222 - 223; texts in square bracktes mine]{treatise2}.
\end{quote}
Notice \label{ambiguo} the ambiguous use of the same name `electromotive force' for denoting two distinct theoretical terms, as already pointed out in section \ref{names}.
\par
Maxwell stresses that, as far as equation (\ref{leggegenmax}) is concerned:
\begin{itemize}
  \item The first term $\vec v\times \vec B$ is due to ``the motion of the particle through the magnetic field \cite[\S 599, p. 223]{treatise2}'';
  \item The second term $-\partial \vec A/\partial t$ ``depends on the time variation
of the magnetic field. This may be due either to the
time - variation of the electric current in the primary circuit, or to
motion of the primary circuit \cite[\S 599, p. 223]{treatise2}''.
  \item The third term $\nabla\varphi$ is introduced ``for the
sake of giving generality to the expression for $\vec E_v$. It
disappears from the integral when extended round the closed circuit.
The quantity $\varphi$ is therefore indeterminate as far as regards the
problem now before us, in which the total electromotive force round
the circuit is to be determined. We shall find, however, that when
we know all the circumstances of the problem, we can assign a
definite value to $\varphi$, and that it represents, according to a certain
definition, the electric potential at the point ($x, y, z$) \cite[\S 598, p. 222]{treatise2}''.
\end{itemize}
In a completely modern form, we write equation (\ref{leggegenmax}) as:
\begin{equation}\label{leggegenmaxmod}
    \mathcal E = \oint_l [\vec E+(\vec v\times \vec B)]\cdot\vec{dl}; \qquad \vec E = -\nabla\varphi -\frac{\partial \vec A}{\partial t}
\end{equation}
This means that the induced {\em emf} in a filiform circuit is the line integral of the Lorentz force on the unit positive charge.
\par
In commenting  his general law (\ref{leggegenmax}), Maxwell speaks of the ``electromotive force [electric field] at a point, or on a particle'': we must intend these words as referring to  ``the
force which would be experienced by a unit of positive electricity
placed at that point''. Therefore, when he speaks of the ``motion of the particle in the magnetic field'', we are leaned to intend the `motion of the unit of positive electricity' or `the motion of the charge'. Anyway,  the velocity which appears in Maxwell's deduction of equation (\ref{leggegenmax}) is the velocity of the circuit element $dl$: an  interpretation in terms of the velocity of the charge presupposes a model of the electric current which Maxwell did not posses.
\section{Electromagnetic induction within Maxwell - Lorentz - Einstein Electromagnetism}\label{mle}

In an axiomatic presentation of MLE electromagnetism, we  begin with the set of Maxwell's equations in vacuum (in the form given to them by Oliver Heaviside)\label{maxeqs}:
\begin{eqnarray}
\nabla \cdot \vec E & = & {{\rho} \over {\varepsilon_0 }}
\label{dive}\\
\nabla \times  \vec E & = & -{{\partial \vec B} \over {\partial
t}} \label{rote}\\
\nabla \cdot \vec B & = & 0 \label{divb}\\
\nabla \times  \vec B & = & \mu_0  \left(\vec J + \varepsilon_0 {{\partial
\vec E} \over {\partial t}}    \right) \label{rotb}
\end{eqnarray}
 where: $\rho(x,y,z,t)$ is a charge density  defined in a finite region of space, $\vec v$ is  the velocity  of $\rho$, $\vec J=\rho\vec v$ and $\varepsilon_0, \mu_0$  are constants to be specified. This set of equations allows to derive the values of $\vec E$ and $\vec B$, named as the electric and the magnetic field, respectively. In fact, a theorem by Helmoltz assures that a vector field is specified if we know its divergence and its curl.
 However, up to now, the electric and the magnetic field are only mathematical symbols with a name: we must attribute physical dimensions to them. We can do so, by stating that, considered a point charge $q$, the fields exert a force on the charge given by:
 \begin{equation}\label{lorforce}
    \vec F= q(\vec E+ \vec v\times \vec B)
     \end{equation}
 This assumption   establishes  the physical dimensions of the two fields, together with those of the two constants $\varepsilon_0, \mu_0$. Equation (\ref{lorforce}), is the expression of the so - called Lorentz force and leads to new predictions. If the charge is at rest, the Lorentz force reduces to  $\vec F = q \vec E$. This last equation is typical of the approach that starts with the Coulomb interaction between two charges at rest. Then, by applying the relativistic transformation equations for the fields and  the forces we get the Lorentz force (\ref{lorforce}).
\par
The assumption of the Lorentz force (\ref{lorforce}) implies that the definition
 of the induced electromotive force in a {\em whatever} closed conducting loop $l$ must be  \cite{ggar}, \cite[pp. 317 - 344, online vers.]{ggib}, \cite{epl}:
\begin{equation}\label{natural}
    \mathcal E = \oint_l (\vec E + \vec v_c \times \vec B)\cdot \vec{dl}
\end{equation}
where $\vec v_c$ is the velocity of a  unit positive point charge and $\vec E, \vec B$ are solutions of Maxwell's equations. This integral yields - numerically -  the work done by the electromagnetic field on a unit positive charge through the entire loop.
The {\em emf} defined by this equation may be considered as the translation of  Maxwell's formula (\ref{leggegenmax}) into MLE electromagnetism. This translation  not only  specifies that the velocity appearing in the formula is the velocity of a unit  positive  point charge but  also grounds the definition of the induced {\em emf} on the assumption of the Lorentz force.
Since:
\begin{equation}\label{vecpot2}
    \vec E = - grad\, \varphi -{{\partial \vec A}\over{\partial t}}
\end{equation}
($\varphi$ scalar potential; $\vec A$ vector potential), equation (\ref{natural}) assumes the form:
\begin{equation}\label{leggegen}
    \mathcal E= -\oint_l \frac{\partial \vec A}{\partial t}\cdot \vec
dl
    +\oint_l (\vec v_{c}\times \vec B)\cdot\vec dl
\end{equation}
 This equation  implies that there are two independent contributions to the induced {\em emf}: the time variation of the vector potential  and
  the effect of the magnetic field on  moving charges.
  The terms $\partial \vec A/\partial t$ and $\vec v_c \times \vec B$ have the dimensions of an electric field. This remark is trivial. However, the two terms imply that an electric field is created at every point of the integration line and that this electric field is responsible for the electric current that circulates in the circuit.
This physical property makes  (\ref{leggegen})  a local law:   it relates the line integral quantity $\mathcal E$ at the instant $t$ to other physical quantities defined at each point of the integration line at the same instant $t$ \footnote{An equation is local if it relates physical quantities at the same point  and at the same instant, or if it relates physical quantities in two distinct points at two subsequent instants $t_1, t_2$, provided that the distance $d$ between the two points satisfies the equation: $d\le c(t_2-t_1)$. This locality condition is a necessary, but not sufficient prerequisite for interpreting causally an equation. }. If the integration line coincides with a rigid, filiform circuit, equation (\ref{natural}) is Lorentz invariant, as it can be easily seen (\ref{inertial}). By the way, a proper description of the inertial relative motion of a magnet and a rigid conducting loop has been considered by Einstein as one of the reasons for the foundation of special relativity \cite[p. 140, eng. trans.]{ein05r}.
  Here, we only stress that, while in the reference frame of the magnet is operative the contribution of the charge moving in the magnetic field, in the reference frame of the loop, the operative term is that due to the time variation of the vector potential. However, both reference frames apply the same equation: this is a typical relativistic feature.
 \par
 If  every element of the circuit is {\em at rest}, $\vec v_c=\vec v_d$, where $\vec v_d$ is the drift velocity of the charge. (Needless to say, this specification was impossible for Maxwell, because he did not have a microscopic model for the electrical current). Then, equation (\ref{leggegen}) assumes the form:
 \begin{equation}\label{leggedrift}
    \mathcal E= -\oint_l \frac{\partial \vec A}{\partial t}\cdot \vec
dl
    +\oint_l (\vec v_{d}\times \vec B)\cdot\vec {dl}
\end{equation}
This equation shows that, in general, the drift velocity contributes to the induced {\em emf}. In filiform circuits, the second line integral is null, because, in every line element,  $\vec v_d$ is parallel to $\vec {dl}$. However, in extended conductors, the drift velocity plays a fundamental role. Particular interesting is the case of Corbino's disc, where the application of equation (\ref{leggedrift}) yields the magneto - resistance effect without the need of using microscopic models  \cite[pp. 3 - 5]{epl}.
 \subsection{Why the  ``flux rule'' is not a physical law}\label{regola}
  The expressions of the induced {\em emf} derived in the previous section contains the vector potential, as in Maxwell's formula.
  However, following the tradition, the induced {\em emf} can be written also in terms of the magnetic field. Starting again from equation (\ref{natural}), we write, in the reference frame of the laboratory:
    \begin{eqnarray}\label{onlyone}
  \mathcal E=  \oint _{l}^{}{\vec E\,\cdot\, \vec{dl}} + \oint_l (\vec v_c\times\vec B) \cdot \vec{dl}&= &\int_{S} \nabla \times \vec E \,
\cdot \, \hat n \, dS + \oint_l (\vec v_c\times\vec B) \cdot \vec{dl}\nonumber\\
&&\\
&= &- \int_{S} {{\partial \vec B} \over
{\partial t}} \, \cdot \, \hat n \, dS + \oint_l (\vec v_c\times\vec B) \cdot \vec{dl}\nonumber
  \end{eqnarray}
  where $S$ is any {\em arbitrary} surface that has the integration line $l$ as contour. Of course, also this equation, like equation (\ref{leggegen}), shows that there are two contributions to the induced {\em emf}: the time variation of the magnetic field and the motion of the charges in the magnetic field.
  Now,  we must use the identity (valid for every vector field with null divergence)  \cite[pp. 323 - 324, online vers.]{ggib}:
  \begin{equation}\label{identity}
  \int_S \frac{\partial \vec B}{\partial t}\cdot  \hat n \,dS=\frac{d}{dt}\int_S \vec B \cdot \hat n \,dS+\oint_l (\vec v_l\times \vec B)\cdot\vec{dl}
  \end{equation}
  where $\vec v_l$,  the velocity of the line element $dl$,  can be different for each line element.
  Then,  equation (\ref{onlyone}) becomes:
  \begin{equation}\label{quasiflusso}
      {\mathcal E}=- \frac{d\Phi}{dt}- \oint_l(\vec v_l\times \vec B)\cdot\vec{dl} +\oint_l (\vec v_c \times \vec B)\cdot \vec{dl}
\end{equation}
In the case of a {\em rigid, filiform  circuit} moving with velocity $V$ along the positive direction of the common $x'\equiv x$ axis, this equation becomes:
\begin{equation}\label{quasiflussorigido}
      {\mathcal E}=- \frac{d\Phi}{dt}- \oint_l(\vec V\times \vec B)\cdot\vec{dl} +\oint_l (\vec v_c \times \vec B)\cdot \vec{dl}
\end{equation}
Since $ V\ll c$ and $ v_d\ll c$, we  put
$\vec v_c= \vec V +\vec v_d$. This last equation with the equal sign instead of the $\approx$ is valid only if  $c=\infty$. Of course, this position  implies the loss of Lorentz invariance.
   Then, finally:
\begin{equation}\label{quasiflux2}
    \mathcal E=-\frac{d\Phi}{dt}
    +\oint_l (\vec v_{d}\times \vec B)\cdot\vec dl= -\frac{d\Phi}{dt}
 \end{equation}
i.e. the ``flux rule''.
 \par
 On the other hand,  in the reference frame of the circuit, we have:
\begin{equation}\label{emirelcirc}
    \mathcal E'= - \int_{S'} {{\partial \vec B'} \over
{\partial t'}} \, \cdot \, \hat n' \, dS' + \oint_{l'} (\vec v'_d\times\vec B') \cdot \vec{dl'}=- \frac{d}{dt'}\int_{S'} \vec B' \cdot\hat n' dS'=-\frac{d\Phi'}{dt'}
\end{equation}
 Hence, the ``flux rule'' is valid in both reference frames -- laboratory's and circuit's -- only if one uses the Galilean transformation of velocities ($c=\infty$).  This suggests that the flux rule -- for rigid and filiform circuits -- may be invariant under a Galilean transformation of coordinates. As a matter of fact, if $c=\infty$,  the magnetic field has the same value in every inertial frame and we see at a glance that the ``flux rule'' is invariant under a Galilean transformation of coordinates \footnote{As pointed out in \cite{levy}, there are two Galilean limits of Electromagnetism. For some more details, see on page \pageref{levy}. }. It is worth adding that, if the circuit is at rest but not  filiform, we get, instead of equation (\ref{quasiflux2}) the equation:
\begin{equation}\label{quasiflux}
    \mathcal E=-\frac{d\Phi}{dt}  +\oint_l (\vec v_{d}\times \vec B)\cdot\vec {dl}
 \end{equation}
Up to now, we have just performed mathematical calculations. Their physical interpretation is as follows.
The ``flux rule'' does not satisfy the locality condition. In fact it relates the {\em emf} induced in a conducting filiform circuit at the instant $t$ to the time variation -- at the same instant $t$ -- of the flux of the magnetic field through an {\em arbitrary} surface that has the circuit as contour. As a consequence, the ``flux rule'' can not be interpreted causally  because what happens at the surface  at the instant $t$ can not influence what happens at the circuit at the same instant $t$, unless physical interactions can propagate with infinite speed. Moreover, since the integration surface can be arbitrarily chosen among those that have the circuit as contour, we should have infinite causes of the same effect.
\par
All the above considerations apart, a comparison between the deduction based on the vector potential and that based on the magnetic field is impressive. The former requires only one passage; the latter is long, cumbersome and full of conceptual pitfalls. Sometimes (often?),  historical developments go through bumpy paths.
\par
The ``flux rule'' not always   yields the correct prediction.   In 1914, Andr\'e Blondel showed that there can be a flux variation without induced {\em emf}.
  Blondel used a solenoid rolled on a wooden cylinder placed between the circular ending plates of an electromagnet. An end of the coil was connected to another parallel wooden cylinder, placed outside the electromagnet (i.e. in a null magnetic field), in such a way that the coil could be unrolled by maintaining the unrolling wire tangent to the two cylinders  \cite{blondel}, \cite[p. 873 - 876]{ggvp}. Blondel performed several experiments; the one which interests us here is that in which the unrolling of the coil -- while producing the variation of the flux of the magnetic field trough the solenoid placed between the plates of the electromagnet -- does not produce any induced {\em emf}. This experiment shows that the ``the flux rule'' is only a calculation tool that must be handled with care and not a physical law.
   The ``flux rule'' can not say where the induced {\em emf} is localized (section \ref{dove?});  in many cases, it yields the correct answer only by choosing {\em ad hoc} the integration line \cite[pp. 704 - 708]{scanlon}, \cite[p. 3]{epl};
it can not describe phenomena in which the drift velocity plays an essential role.
\par
\begin{figure}[h]
  \centering
  \includegraphics[width=3cm]{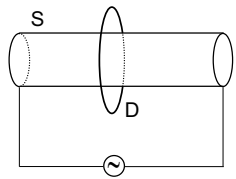}\\
  \caption{$S$ is a long solenoid of radius $a$ in which  flows a low frequency alternate current $I$. In the ring $D$ of radius $R$, placed symmetrically in a plane perpendicular to the solenoid axis, an {\em emf} is induced.}\label{ml}
\end{figure}

The fact that the ``flux rule'' is only a calculation tool can be  illustrated also   by an experiment in which the induced circuit is placed in a region of space where there is a  vector potential field but no magnetic field  (fig. \ref{ml}). Though this experiment does not yield new support to  the epistemological stand attributed  to the ``flux rule'' in the present paper, it is interesting because it can be considered the classical analog of the Aharonov - Bohm effect and it is denoted as the Maxwell - Lodge effect \cite{rousseaux}.
\par\noindent
 In the ideal case of an infinitely long solenoid,  the magnetic field
outside the solenoid is zero. According to the general law of induction (\ref{leggegen}), the {\em emf} induced in the filiform ring is given by:
\begin{equation}\label{anello}
    \mathcal E = -\oint_{ring} \frac{\partial \vec A}{\partial t} \cdot
\vec{dl}=
    -\frac{d}{dt}\oint_{ring} \vec A\cdot \vec {dl}= - 2\pi R \frac{dA}{dt}
\end{equation}
where $R$ is the radius of the ring. If $dA/dt>0$, $\mathcal E<0$: this means that the current in the ring circulates in the opposite direction of the current in the solenoid. The values of the vector potential can be calculated directly from the known value of the current in the solenoid \cite{rousseaux}.
Alternatively, the induced {\em emf} (\ref{anello}) can be calculated more easily  by applying the ``flux rule'':
 \begin{equation}\label{anello_regola}
    \mathcal E =-\frac{d}{dt}\int_S \vec B\cdot\hat n\, dS=-\mu_0n\pi a^2\frac{dI}{dt}
 \end{equation}
 where $S$ is an arbitrary surface that has the ring as a contour and $n$ is the number of turns per unit length of the solenoid.
In the ideal case, the induced {\em emf} does not depend on the radius $R$ of the ring.
 The vector potential outside the solenoid is a vector tangent to a circumference centered on the solenoid axis and directed as the current in the solenoid. Its value can be calculated
  by equating the last members of the  two equations above:
 \begin{equation}\label{potvett}
    A=\mu_0\frac{n a^2 }{2R}I
 \end{equation}
 where the constant of integration has been put equal to zero, in order to obtain a vector field that vanishes  far away from the source.
 Also in this case, the ``flux rule'' relates what happens in the ring at an instant to what happens at the same instant  in the portion of the integration surface $S$ which intersects the solenoid.   Also in this case, this relation can not be interpreted as a causal one because no physical interaction can propagate with infinite speed.
 If one takes into account the fact that the length of the solenoid is finite, the magnetic field outside the solenoid is not null. This implies that the value of the vector potential is smaller than that calculated by equation (\ref{potvett}) and, consequently,  the measured {\em emf} is smaller than that given  by equation (\ref{anello}) \cite[pp. 253 - 254]{rousseaux}.
\subsection{The basic importance of a correct definition of the induced {\em emf}}\label{nota}
Often, instead of (\ref{natural}),  the induced {\em emf} is defined as:
\begin{equation}\label{emiC}
    \mathcal E_C=  \oint_l\vec E\cdot \vec {dl}
\end{equation}
This definition (`C' stays for Coulomb) follows from the definition  of the force on a charge at rest:
\begin{equation}\label{coulomb}
     \vec F= q\vec E
\end{equation}
While equation (\ref{coulomb}) is correct, the definition (\ref{emiC}), if improperly used,  leads to wrong predictions.
  In fact, the definition (\ref{emiC}) can be used only in the reference frame of a {\em rigid and filiform}  circuit. In particular, it can not be applied when the circuit or part of it is in motion.
 \par
Since:
\begin{equation}\label{eqmax2}
    \nabla \times {\vec E}=-\frac{\partial \vec B}{\partial t}
\end{equation}
it follows that:
\begin{equation}\label{intele}
  \mathcal E_C=   \oint_l\vec E\cdot \vec {dl}=\oint_s \nabla\times {\vec E}\cdot \hat n \,dS=-\oint_S \frac{\partial \vec B}{\partial t}\cdot \hat n  \,dS
\end{equation}
where $S$ is any (arbitrary) surface that has the integration line $l$ as contour.
If the integration line does not change with time (circuit at rest), we get:
\begin{equation}\label{flussocou}
  \mathcal E_C=   -\frac{d}{dt}\int_S \vec B\cdot\hat n\,dS=-\frac{d\Phi}{dt}
  \end{equation}
  i.e. the ``flux rule''.
  At this point, one might be tempted to ask what happens if we let the circuit (or part of it) move. Then, by re - starting from equation
  (\ref{intele}) and by using equation (\ref{identity}), one gets immediately:
  \begin{equation}\label{noflux}
         {\mathcal E_C}=-
{{d}\over{dt}}\int_{S} {\vec B \, \cdot \, \hat n \, dS}- \oint_l(\vec v_l\times \vec B)\cdot\vec{dl} =-\frac{d\Phi}{dt}- \oint_l(\vec v_l\times \vec B)\cdot\vec{dl}
\end{equation}
 This equation is wrong, the correct one being equation (\ref{quasiflux}). The reason is that the definition of the induced {\em emf} (\ref{emiC}) lacks the term
 \begin{equation}\label{manca}
    \oint_l (\vec v_c\times\vec B)\cdot \vec {dl}
 \end{equation}
 coming from the second term of the correct definition of induced {\em emf} (\ref{natural}). If we add this term, equation (\ref{noflux}) transforms into the correct one (\ref{quasiflux}) or in its approximation for low velocities (\ref{quasiflux2}).
 \par
 The above considerations will be useful in the discussion of how electromagnetic induction is treated in  textbooks (section \ref{why}).
 \subsection{Where is localized the induced {\em emf}?}\label{dove?}
\begin{figure}[h]
\centerline{
\includegraphics[width=4cm]{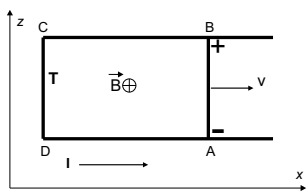}
}
 \caption{\label{barra}
 The  filiform, rigid wire AB of length $a$ moves with constant velocity $\vec v$ along the U - shaped frame T in a uniform and constant magnetic field $\vec B$ -- produced by a source at rest in the reference frame of T -- and directed along the positive direction of the $y$ axis.}
\end{figure}

 Einstein, among others, held that this question is meaningless: ``Questions as to the `seat' of electrodynamic electromotive
forces (unipolar machines) also becomes pointless \cite[p. 159, eng. transl.]{ein05r}.'' Instead, we shall show that this  question has physical meaning, in the sense discussed in section \ref{meaning}, by considering  the illustrative example of a  rigid wire moving along an U shaped conducting frame (fig. \ref{barra}).
  \par\noindent
This thought  experiment  goes back to Maxwell \cite[pp. 218 - 219, \S\, 594]{treatise2}. Conceptually, it can be considered as a version  of Faraday's disc in which the rotation of the disc is substituted by the inertial motion of a part of the circuit, thus allowing the use of two inertial reference frames.
According to the general law (\ref{leggegen}), the induced {\em emf} is due to the magnetic component of the Lorentz force and  is given by $\mathcal E= B_yva$: the {\em emf} is localized in the bar, as it can be shown  \cite[pp. 325 - 330, online vers.]{ggib}. Here, we recall only the results. The bar acts as a battery, with the positive pole at the point $B$. Experimentally, this implies that, in the wire AB, the current circulates from  $A$ to $B$, i.e. from the point at lower potential to that at higher potential, like in a battery. Instead, taken a pair of points on the frame, the current circulates from the point at higher potential to that at lower potential. The concept of  localization of the induced {\em emf} has physical meaning because it allows the prediction -- testable by experiment --  of how the potential difference between two points is related to the  direction of circulation of the current.  If we drop this concept, we decrease the predictive power of the theory.
 \par
 As it can be easily seen, the flux rule predicts the correct induced {\em emf}. The flux of the magnetic field through the area ABCD is given by $\Phi= B_yavt$, if at the instant $t=0$ AB coincides with DC and the circulation direction is clockwise. Then $\mathcal E= -B_yav$, where the minus sign indicates that the induced current circulates counterclockwise, as indicated in the figure.   The ``flux rule''  can not say where the {\em emf} is localized; it can only guess that it might  be localized in the wire AB because it is moving (on the ground  that it is the wire's movement that produces the variation of the magnetic field flux  and, hence, the induced {\em emf}). However, in the reference frame of AB, the ``flux rule'' will say that the {\em emf} is localized in the opposite vertical arm CD of the frame, because it is moving: this statement is false, because, also in the reference frame of AB, the {\em emf} is localized in  AB.
In fact, in every point of the AB's frame, there is an electric field -- due to the fields' transformation equations --  given by:
\begin{equation}\label{elezeta}
  E_z'=\Gamma  vB_y
\end{equation}
Therefore, in AB, an {\em emf} $\mathcal E'=\Gamma vB_y a=\Gamma \mathcal E$ is induced. On the other hand, in CD, the electric field $E_z'$ is exactly balanced by the magnetic component of the Lorentz force.
The relation $\mathcal E'=\Gamma\mathcal E$ is a particular case of the more general one treated in  \ref{inertial}.
Of course, for maintaining the wire AB in motion with constant velocity, a mechanical power must be supplied. It is immediately found that this power is equal to the electrical power dissipated in the circuit.
 \subsection{The unipolar induction}\label{faruni}
\begin{figure}[h]
\centerline{
 \includegraphics[width=4cm]{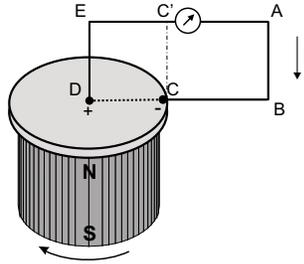}
} \caption{\label{far_ila}  Faraday's disc. A conducting disc rotates about its axis with constant angular velocity $\Omega$ in the clockwise direction. The disc is immersed in a uniform magnetic field generated by the cylindrical magnet, electrically isolated from the disc. The magnetic field at the disc is pointing upward.  C and D are two sliding  contacts. A current $I$ flows in the circuit DEABCD. The radius DC acts as a battery. The same results are obtained if the magnet rotates with the disc.}
\end{figure}

In spite of Faraday's statement that the so called unipolar induction is only a particular case of Faraday's disc, there is a rather ample literature which holds that the unipolar induction is still a problematic case (see sections \ref{unisec}, \ref{why}). For this reason, we shall treat here in some detail the case of Faraday's disc (fig \ref{far_ila}) and its unipolar version.
The Faraday's disc is easily treated by using the general law of electromagnetic induction (\ref{leggegen}).
\par\noindent
  If we neglect the drift velocity of the charges in the disc, the general law (\ref{leggegen})  yields for the induced {\em emf} (the vector potential is constant):
  \begin{equation}\label{fem_faraday}
{\mathcal E} = \int_{0}^{a}{\Omega r B dr}=  {{1} \over {2}} \Omega B a^2
\end{equation}
where $a$ is the radius of the disc. The current $I$ is then given by $I= \mathcal E/R$ where $R$ is the electrical resistance of the entire circuit.
The {\em emf} is localized in the radius CD of the disc, with $V_D>V_C$. Then, the radius CD acts as a battery. The  current in it circulates from the point   at lower potential C to that at higher potential D. Instead,  taken a pair of points in the external circuit, the current circulates from the point at higher potential to that at lower potential, as it should be in an Ohmic element.
This treatment is a  quantitative one and allows the experimental verification of all the physical quantities involved. Exactly the same result is obtained if the magnet rotates with the disc. On the other hand, if only the magnet rotates, the induced {\em emf} is null, since is null each of the two line integrals of the general law.
\par
\begin{figure}[h]
  \centering{
  \includegraphics[width=2.0cm]{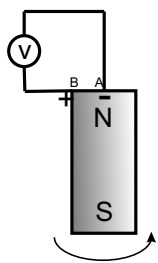}\\
      }
  \caption{Unipolar induction. The cylindrical, conducting magnet NS rotates about its axis with angular velocity $\Omega$. An {\em emf} is induced in the radius $AB$ of length $a$, and its value is given by $\mathcal E = (1/2) \Omega B a^2$.}\label{unipolar}
\end{figure}

The fact that the unipolar induction is only and simply a particular case of the Faraday's disc, can be seen by considering  figure \ref{unipolar}.
\par\noindent
In this arrangement, the upper surface of the conducting magnet plays the role of the copper disc in the Faraday's disc arrangement.
If we move the sliding contact $B$ to a point on the lateral surface of the magnet, nothing changes from the conceptual point of view.
If we take into account the fact that the magnet is an extended conductor, and -- consequently -- we take into account the drift velocity of the charges, the calculations become much more complex, as shown in \cite[p. 4]{epl} for the simple case of Faraday's disc with circular symmetry.
\par
\begin{figure}[h]
  \centering{
  \includegraphics[width=2.5cm]{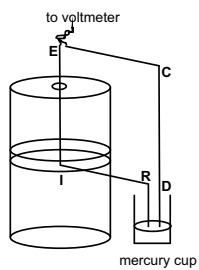}\\
  }
  \caption{A cylindrical magnet is made up by a stack of ceramic ring  magnets: an equatorial gap allows the use of the loop $EI+IR+ECD$. $IR$ and $ECD$ are connected through a mercury cup. The cylindrical magnet and the circuits $IR$ and $ECD$ can perform, independently or together, small rotational oscillations.   The system is enclosed in a ferromagnetic frame (not shown in the figure) that essentially shields the circuit $ECD$ from the magnetic field of the cylindrical magnet. Therefore, no contribution to the measured {\em emf} comes from the circuit $ECD$ \cite{muller}.}\label{muller}
\end{figure}

Recently, some basic Faraday's experiments has been reproduced by M\"{u}ller with an arrangement that avoids the complications due to extended materials \cite{muller}. The cylindrical magnet is made up by a stack of ceramic ring  magnets: an equatorial gap allows the use of a rectangular loop, as shown in figure \ref{muller}. The setup allows to swing linearly or rotationally, singularly or together, several parts of the apparatus. In this way, many Faraday's experiments can be reproduced with only one apparatus.
\par\noindent
Let us see how.
 With small rotational oscillations of $IR$ or $IR\, +\, magnet$  or of the magnet, all experiments shown in  table \ref{discot}  are carried out.
  \begin{table}[h]\small
\begin{center}\leavevmode
\begin{tabular}{|l|c|c|}
\hline \bf{What}&\bf{Relative motion}
& \bf{Induced}\\
\bf  is rotating?& \bf{wire IR~-~magnet}& \bf{current}\\
\hline\hline
I. Wire IR &Yes  &Yes \\
\hline
II. Magnet&Yes &No\\
\hline
III. Wire IR and magnet&No &Yes \\
\hline
\end{tabular}
\caption{Phenomena that can observed with  M\"{u}ller's apparatus (fig. \ref{muller}). The reference frame is that of the
laboratory.\label{discot}}
\end{center}
\end{table}
\par\noindent
 In particular:
 \begin{itemize}
   \item rotational oscillations of $IR$ correspond to experiment I of table \ref{discot}
   \item rotational oscillations of the magnet correspond to experiment II
   \item rotational oscillations of $IR$ {\em and}  of the magnet correspond to experiment III and, at the same time, to the unipolar induction experiment (since $IR$ is ``embedded'' into the magnet)
 \end{itemize}
\label{non2}  The experimental results confirm all Faraday's ones carried out with the disc or the conducting magnet (unipolar induction).
  In particular, the experiment with only the magnet in oscillation (experiment II of table \ref{discot}) shows that the lines of magnetic force {\em do not rotate}, as Faraday  showed about two centuries before.
Quantitatively, the {\em emf} induced in the radius $IR$ is given by:
\begin{equation}\label{radius}
    \mathcal E= \frac{1}{2}B\Omega(R_2^2-R_1^2)
\end{equation}
where $B$ is some average value of the magnetic field in the gap containing $IR$.
The measured and the theoretical values are in reasonable agreement. It must be recalled that also Blondel performed an experiment conceptually equivalent to the Faraday's disc and verified the validity of equation (\ref{radius}). Precisely, it was experiment IV discussed in \cite[pp. 89 - 90]{blondel} and in \cite[p. 874]{ggvp}.
Finally, also the experiments with the various components oscillating linearly, confirm Faraday's fundamental distinction between linear and rotational relative motion of magnet and conductor.
\subsection{About lines of magnetic force that {\em (do not)} move and {\em (do not)} rotate}\label{unisec}
 Within MLE electromagnetism, the lines of magnetic force  are only a way of illustrating graphically the spatial variations of the magnetic field.
  Indeed, every  prediction of MLE electromagnetism can be made without using the concept of lines of magnetic force.
  To put it plainly: Maxwell's equations are about fields, not about lines of force.  This is a case in which the criteria  enunciated  in section \ref{meaning} can be usefully employed. According to the strong criterion, the concept of lines of magnetic force had physical meaning in Faraday's theory of electromagnetic induction because its elimination  would have reduced the predictive power of the theory (to zero in this case). Instead, in MLE electromagnetism, the same concept satisfies only the weak criterion. In fact, in MLE electromagnetism, while the concept of lines of magnetic force can be dropped without diminishing the predictive power of the theory, its elimination diminishes the descriptive capacity of the theory.
  \par
 Given this, if somewhere in the literature we encounter expressions like `rotation of lines of magnetic force' we should at once recognize an improper use of  Faraday's concept. Improper because Faraday repeatedly stressed that -- on the basis of his experimental findings --  the lines of magnetic force do not rotate.
\par
    In  three popular textbooks of the first decades of Nineteenth century, written by Eduard Riecke (1845 - 1915), Henri Bouasse (1866 - 1953) and Orest Danilovich Chwolson (1852 - 1934),  it is held that the unipolar induction can be described  {\em also} by supposing that the lines of magnetic force rotate with the magnet \cite[ pp. 213 - 214]{riecke}, \cite[pp. 374 - 377]{bouasse}, \cite[pp. 120 - 122]{russo}. This assertion implies to attribute to the lines  of magnetic force a velocity  different from that of light. This assumption strikingly contrasts with Faraday's view according to which  the changes of lines of magnetic force  propagate with the velocity of light, or instantaneously.  Unexpectedly, the idea of rotating or moving lines of magnetic force has been resumed recently: see, for instance,  \cite{brotante}; or  \cite[p. 6]{terraexp}, where it is considered, in a secondary passage, only  as a possibility not excluded in principle.
  Therefore, it seems worth trying to further clarify this point.
  \par
    If a cylindrical magnet rotates in the laboratory frame with angular velocity $\Omega$, we  assume -- in contrast with Faraday's view and in contrast with MLE electromagnetism --  that at a point whose distance from the magnet's axis is $r$, a velocity $\Omega r$ must be attributed to the field, with the understanding that the field rotates in the same direction of the magnet: the magnet drags its magnetic field in its motion. Of course,   for appropriate values of $\Omega$ and $r$, $\Omega r$ may be greater than $c$. But we keep going ahead.
   Next, we must decide which is the effect of the field moving with velocity $\vec v_B$ on a charge {\em at rest}.
     This means that, instead of the Lorentz force, we could write, for instance:
     \begin{equation}\label{ruotaruota}
        \mathcal F= q[\vec E + (\vec v_c\times \vec B )+ (\vec B\times \vec v_B)]
     \end{equation}
     where $\vec v_c$ is the velocity of the charge and $\vec v_B$ that of the magnetic field. For satisfying the relativity principle, this equation should be valid in every inertial frame.
     Of course, equation  (\ref{ruotaruota}) is incompatible with MLE electromagnetism. For two reasons: because it modifies the fundamental equation which gives physical meaning to the fields; and because it modifies an equation that has been confirmed by innumerable experiments. Equation (\ref{ruotaruota}) has another weird  property: it contains two terms, those in round brackets,  that contradict each other. The former  says that the magnetic field exerts  no action on a charge  at rest; the latter says that the magnetic field does. It seems as a sleight of hand. Equation (\ref{ruotaruota}) is never written by the supporters of this heterodox idea, though they use, at will,  some pieces of it.
       \par
       We shall see how equation (\ref{ruotaruota}) works in two exemplar cases. Firstly, let  us consider -- in the laboratory reference frame -- a positive charge $q$ that is moving with velocity $\vec v_c$ along the  positive direction of the $x$ axis, in   a uniform and constant  magnetic field directed along the positive direction of the $z$ axis; the magnetic field is generated by a source at rest in the laboratory. According to the formula (\ref{ruotaruota}), on the charge is exerted a   force given by $\mathcal F= q(\vec v_c\times\vec B)$ which has only the component $\mathcal F_y=-qv_cB_z$ different from zero. This force is due to first term in round brackets;  the contribution of the second term is null, since, in the reference frame of the magnet the magnetic field is at rest.
       In the inertial reference frame co - moving with the charge, the source of the magnetic field {\em and} the magnetic field move with velocity $\vec v'_B=- \vec v_c$.
   Then, on the charge is exerted a force  given by:
    \begin{equation}\label{ruota}
        \mathcal F'= q[\vec E' +  (\vec B'\times \vec v'_B)]=  q[\vec E' -  (\vec B'\times \vec v_c)]= q[\vec E' +  (\vec v_c\times \vec B')]
     \end{equation}
   that -- taking into account the fields transformations --  has only the $y$ component different from zero and this is given by $\mathcal F'_y=-\Gamma v_cB_z-\Gamma v_c B_z=- 2\Gamma v_cB_z$. This force is twice  the correct one.  Therefore, as expected, the assumption (\ref{ruotaruota}) is incompatible with MLE electromagnetism. Of course, if we drop the second term in the round brackets in equation (\ref{ruotaruota}), we come back to the description of MLE electromagnetism. Instead, if we drop the first term in round brackets (i.e. the magnetic component of the Lorentz force) we see that the hypothesis of the rotating magnetic field mimics the correct result. However, this is accomplished at the cost of applying equation (\ref{ruotaruota}) only in the reference frame of the moving charge, i.e. at the cost of  violating the relativity principle. Therefore, the hypothesis of rotating magnetic field is without any theoretical foundation.
\par
As a second example, let us consider
 the case of  Faraday's disc in the configuration in which the magnet rotates with the disc (fig. \ref{far_ila}). As we have seen, the unipolar induction is only a particular case of Faraday's disc. So, what is valid for Faraday's disc is valid also for its unipolar variant, considered by Riecke, Bouasse and Chwolson. According to MLE electromagnetism, there is an induced {\em emf} that is localized in the radius CD with $V_D>V_C$. Instead, if we consider the alternative description based {\em only} on the second term in round brackets of equation (\ref{ruotaruota}), we can proceed in the following way. We suppose
 AE to be very close to  BC: then the value of the magnetic field and of its  velocity will be approximately the same in every pair of corresponding points of  AC$'$ and BC. As a consequence, their contributions to the induced {\em emf} will be the same but  opposite to each other. The contribution from   ED is null, since the magnetic field is at rest in every point and that from AB can be neglected because its length can be reduced at will. Hence, we are left with the contributions coming from CD and C$'$E. The contribution coming from CD is null, since the charges in it are not at rest. Therefore, there is only the contribution coming from C$'$E:  C$'$E acts as a battery with the positive pole at C$'$: hence the circulation of the current is clockwise, as it must, and  the induced {\em emf} is approximately equal to that calculated by considering the magnetic component of the Lorentz force operating in the radius CD of the disc. In this way, we  see how -- {\em in the limit considered} --  the idea of rotating magnetic field can predict the correct value of the induced {\em emf} and the correct direction of the induced current. Of course, as we know, a simple measurement discriminates between the two descriptions: the {\em emf} is localized in the radius CD of the disc and not in the segment C$'$E.  Instead, if we keep both terms in round brackets, the predicted {\em emf} will be twice the correct one. Let us now consider the case in which only the magnet rotates. In this case, let us assume that AB coincides with CC$'$ an C$'$E is far away from CD. Then, the contributions to the induced {\em emf} due to the second term in round brackets of equation (\ref{ruotaruota}) from the various segments are: zero from DE, because in every point $\vec v_B=0$, zero from CC$'$, because the vector $\vec B\times\vec v_B$ is perpendicular to the wire CC$'$; zero from C$'$E because the magnetic field is vanishingly small; equal to $(1/2)B\Omega a^2$ from CD. This prediction is falsified by the experiment of Faraday and M\"{u}ller:  there is no induced {\em emf}. As already noticed, the null result of this experiment is predicted by the general law (\ref{leggegen}): each of its two line integrals is null. \par
If the appearance of the idea of rotating lines of magnetic force in the first decades of the twentieth century can be figured out in a period in which MLE electromagnetism was still to be completely assimilated,  its nowadays reappearance is hard to comprehend \cite{brotante}. The aim of the paper by Leus and Taylor is that of finding an experimental support -- support, not proof -- of the hypothesis of rotating magnetic field. The experimental setup is conceptually identical to that of Faraday's disc with the magnet co - rotating with the disc, but with the  drawback that  the values  of the magnetic field in a region in which there is a   part of the rotating circuit are not known.
These missing data impede  a reliable quantitative prediction of the experimental results, unless one is intended to ignore the magnetic component of the Lorentz force, as the authors do.
In the introduction of the paper, we  find a standard formula  according to which there are two contributions to  the induced {\em emf}: the time variation of the magnetic field and the movement of the charges in the magnetic field. This last term contains, obviously, the magnetic component of the Lorentz force. However, from this point on, the second term in round brackets of equation (\ref{ruotaruota}) is introduced as an independent postulate, without noticing
 its incompatibility  with MLE electromagnetism.
The paper does not -- and could have not -- contain  new experimental results, with respect to those obtained with the Faraday's disc co - rotating with the magnet. The description of the induced {\em emf} is given by using only the second term in round brackets of equation (\ref{ruotaruota}), as it has been done above for the Faraday's disc co - rotating with the magnet, but without considering the alternative description based on the magnetic component of the Lorentz force. Therefore, the authors  missed the only possible experimental test capable of discriminating between the two description: to measure where the induced {\em emf} is localized.
\par
It may be of some utility to summarize the content of this section in few, basic points:
\begin{itemize}
  \item The concept of lines of magnetic force, fundamental for Faraday's theory of electromagnetic induction, is not a basic one within MLE electromagnetism. It serves only as a graphical  illustration tool.
        \item Faraday conceived the lines of magnetic force as independent from matter. Therefore, the lines of magnetic force do not participate in the motion of matter. Instead, their changes propagate with the velocity of light or instantaneously.
  \item The hypothesis  of attributing to the magnetic field a velocity typical of a material entity is completely extraneous to Faraday's theory of electromagnetic induction and it is incompatible with MLE electromagnetism.
           \item This hypothesis is falsified by experiment.
\end{itemize}
   \section{Why Maxwell's general law has been forgotten?}\label{why}
 We shall try to understand why Maxwell's {\em general law} of electromagnetic induction has been forgotten and why, instead, the ``flux rule'' has  taken root. Before all, it must be stressed that Maxwell, after having deduced his general law, did not made any comment about the relation between the general law and the ``flux rule''.  This lacking comment, has  probably not favored  the acceptance of the general law, because the ``flux rule'' and the general law could have been considered by an inattentive reader as somehow equivalent.
 It has been suggested that the elimination of the vector potential by Hertz and Heaviside from the presentation of Maxwell's Electromagnetism could have contributed to the oblivion of Maxwell's general law \cite[pp. 871 - 872]{ggvp}, \cite[p. 91]{erq}. This is a plausible  hypothesis. However, other factors may have contributed. To find out these  other  factors, we shall look at some university textbooks, considered as significant on the basis of their authors and/or of their diffusion or popularity.
 \par
 Let us begin by looking at   the first
 decades of Nineteenth century, namely far enough from Maxwell's {\em Treatise}, but only few years after Einstein's reformulation of Maxwell's electromagnetism. In those years,  three textbooks  have reached a wide appreciation and popularity. The authors were Eduard Riecke, Henri Bouasse and Orest Danilovich Chwolson   \footnote{The unpublished working notes of my coworker Paolantonio Marazzini (1940 - 2013) have guided me in the choice of these three texts. Marazzini, in 2009  began a study on the theoretical description of electromagnetic induction from Maxwell's {\em Treatise} to about 1940. As acknowledged in \cite{ggvp}, from this Marazzini's research, I came to know about Blondel's work. }.
 Riecke  presents the phenomena of electromagnetic induction by discussing a series of experiments, most of them taken from Faraday's {\em Researches}. He discusses the phenomena of `magnetic induction', due to the relative motion of a magnet and a conducting loop and those of `electric induction', due to the time variation of the current in a circuit, in two different, not consecutive chapters. The names used are the same as Faraday's.
  The theoretical treatment of the `magnetic induction' is based on the idea that there is induced current in a closed loop if the number of lines of magnetic force across the loop changes. This passage signs the abandon of  Faraday's local theory towards the ``flux rule''. Quantitatively, this statement is specified by saying that ``If a closed loop moves in a magnetic field, the integral value of the electromotive force is proportional to the algebraic difference between the number of lines of force  across the loop at the beginning and at the end of the motion \cite[\S 532, p. 257]{riecke}.''
Nowhere, the ``flux rule'' is written as an equation. The equation of the ``flux rule'' does not appear also in the treatment of the `electric induction'. The first edition of Riecke's textbook (1896) received a  review by Alfred Gray in the pages of {\em Nature} \cite{gray}. Its closing sentence was:
\begin{quote}\small
  In taking leave of this treatise we wish to say that
students owe much to Prof. Riecke for giving them a
readable, not too abstruse, and yet thoroughly sound
and fairly full discussion of the elements of physics. To
many German students who have not time to struggle
through the larger treatises this book must be very
welcome \cite[p. 567]{gray}.
\end{quote}
 Riecke's textbook on Electromagnetism has been reprinted in 1928 \cite{riecke1928}.
\par
The book by Henri Bouasse (1866 - 1953) is  more sophisticated, owing to the  role reserved to the mathematical treatments. It is also peculiar, for  the epistemological  and didactic choices.
The title of the book is: ``{\em Magnetism and Electricity - Part one - Study of the Magnetic Field}''.
Nowhere in the book one can find the word ``electric field'', because:
 \begin{quote}\small
 \dots  in no application, you hear well, in no one,  the electric field steps in. Why talk about it first?
In order to avoid any dispute on this subject, the second volume of this {\em Course on Magnetism and Electricity} is devoted to applications, the third to the electric field. The reader will verify that the word {\em electric field}  is not printed once in the thousand pages contained in the first two volumes \cite[p. XIV; original italics]{bouasse}.
\end{quote}
It is true. However, the word is simply omitted, because the electric field is there in the equations. Throughout the entire volume, the term ``electromotive force'' is used with our meaning: it has the dimensions of an electric potential. However, as in Maxwell's Treatise, the same term denotes what we call ``electric field''.
After having discussed some typical experiments of electromagnetic induction, Bouasse writes:
\begin{quote}\small
A closed circuit (of invariable or variable shape) moves in a [magnetic] field; the variation of the flux of the [magnetic] induction through an arbitrary surface, with the circuit as a contour, creates an electromotive force, whose absolute value in given in volts by the relation:
    \begin{equation}\label{flussobouasse}
        \mathcal E= 10^{-8}\frac{d\Phi}{dt}
    \end{equation}
  $t$ represents the temp. The direction of this electromotive force is such that the current which it tends to produce will be the cause of the electromagnetic forces which oppose the displacement [of the circuit] \cite[p. 275]{bouasse}.
\end{quote}
(I have changed the symbols for uniformity with those used in this paper).
Some  pages ahead, in a paragraph entitled ``Full expression of the electromotive force'' [electric field], Bouasse writes the expression of the electromotive force [electric field] at a point of a circuit as the sum of three terms (in modern symbols) \cite[p. 322 - 323]{bouasse}:
\begin{itemize}
  \item  $-\nabla \varphi$: electrostatic {\em electromotive force} [electric field]
  \item $-\partial\vec A/\partial t$: induced {\em electromotive force} [electric field]
  \item the electromotive force [electric field] of non - homogeneity (as in the batteries)
\end{itemize}
When the circuit is moving, we must add a fourth term $\vec v\times\vec B$ due to the cut flux of the `magnetic induction'. Of course, the total electromotive force acting in a circuit, is the line integral of the sum of the four terms.
 Differently from Maxwell -- who derived the general law in a deductive way within a Lagrangian treatment of the currents --  Bouasse builds up the general law in a constructive way, piece by piece.
Bouasse does not cite Maxwell, coherently with his choice of avoiding any quotation.
\par
Daniel Abramovich Chwolson (1819 - 1911) wrote a physics textbook composed by five volumes. It has been translated in French, German and Spanish.
The treatment of electromagnetic induction appears in the second chapter of the fifth volume,  written by Alexander Antonovich Dobiash (1875 - 1932) \cite{russo}.
The treatment begins with this statement:
\begin{quote}\small
    Consider a wire placed in a
magnetic field. When, for any reason, the number of lines
of magnetic induction which cross the surface limited by this wire comes to
change, a current start flowing in the wire.  Such a current is called
induced current. It ceases as soon as the cause which produced it disappears, that is to say
as soon as the number of magnetic induction lines, which pass through
 its contour, no longer varies \cite[p. 44 - 45]{russo}.
\end{quote}
This is the ``flux  rule''. The way in which it is presented strongly suggests a causal relation between the flux variation and the induced electromotive force.  The following pages  are characterized  by a somewhat blurred discussion of some experiments.
  Mainly, the induced {\em emf} is related to the relative motion between the circuit and the lines of magnetic force, as in Faraday's theory. From this discussion, emerges the formula $\vec v\times \vec B$, just for finding out the direction of the induced {\em emf}  \cite[p. 47]{russo}. The case of a time varying magnetic field is cited, but not treated. At last, the ``flux rule'' is derived -- again -- by considering the motion of a circuit in a magnetic field \cite[p. 49 - 50]{russo}. The chapter, after the treatment of some applications like self - induction, alternate and eddy currents, ends with the discussion of the unipolar induction \cite[p. 120 - 122]{russo}. As already recalled and discussed in section \ref{unisec},
  all three textbooks, in discussing the unipolar induction, hold that it can be explained both by supposing the lines of magnetic force at rest or rotating with the magnet (Riecke, pp. 213 - 214; Bouasse, pp. 374 - 377).  For a historical reconstruction of unipolar induction's issue, see, for instance, \cite{miller}.
     \par
After the second world war, the number of textbooks on Electromagnetism has increased in an impressive way. Therefore, the only viable choice is that of  picking up a few texts, mainly on the basis of their authors and/or their diffusion.
 \par
 The book by John Slater (1900 - 1976)  and Nathaniel Frank (1904 - 1984), entitled {\em Electromagnetism}, was published in 1947.  Electromagnetic induction is treated in three pages, half of them of introductory character. At the beginning of the theoretical treatment, we find a general and careful  definition of the electromotive force:
\begin{quote}\small
    By definition, the {\em emf} around a
circuit equals the total work done, both by electric and magnetic
forces and by any other sort of forces, such as those concerned in
chemical processes, per unit charge, in carrying a charge around the
circuit \cite[p. 79]{sf}.
\end{quote}
However, this  premise does not lead the authors to define the induced {\em emf} as the line integral of the Lorentz force on a unit positive charge as in equation (\ref{natural}). In fact, soon after,
 it is stated that the phenomena of electromagnetic induction are described by ``Faraday's law in integral form'' (I have changed the symbols for uniformity with those used in this paper):
\begin{equation}\label{farlaw}
    \oint \vec E\cdot \vec {dl}=-\frac{d}{dt}\int_S \vec B \cdot \hat n\, dS
\end{equation}
When the circuit is fixed and the magnetic field depends on time, it is straightforward to derive from (\ref{farlaw}), by using Stokes' theorem, the Maxwell's equation:
\begin{equation}\label{terzasf}
    \nabla\times\vec E=-\frac{\partial \vec B}{\partial t}
\end{equation}
The path from the `Faraday's law' (\ref{farlaw}) (considered as an experimental result) to Maxwell's equation (\ref{terzasf}) is typical in many textbooks. Also the reverse path is very common: assumed the validity of Maxwell's equation (\ref{terzasf}), the `Faraday's law' (\ref{farlaw}) is obtained. Clearly, for Slater and Frank, the electromagnetic induction is -- from a teaching viewpoint --  an unproblematic issue, which does not deserve particular attention. This stand seems typical of textbooks that are considered by their authors as an exemplification of Theoretical Physics.
\par
In order to find a critical approach to  electromagnetic induction, we must wait for the second volume of Feynman's {\em Lectures on Physics} \cite{feyn2}.   Chapter XVII is entitled ``The laws of induction'' and is dedicated to the theoretical treatment of the experiments and technical devices presented in the previous chapter (``Induced currents''). In this chapter, the ``flux rule'' is enunciated: the inverted commas are Feynman's. According to Feynman, we are dealing with a rule, not with a physical law. Though Feynman does not use our term of ``calculation tool'', it is the same thing. The reason lies in the fact that there are ``exceptions to the ``flux rule'''', discussed in a special paragraph (17.2). Feynman considers two cases: a version of Faraday's disc and that of the ``rocking plates'' and conclude that the ``flux rule'' does not work in these cases because:
\begin{quote}\label{feynman}\small
{\em It must be applied to circuits in which the
material of the circuit remains the same}. When the material of the circuit is changing,
we must return to the basic laws. The correct physics is always given by the
two basic laws:
\begin{equation}
\nonumber   \vec F = q(\vec E + \vec v\times\vec B)
\end{equation}
\begin{equation}
  \nabla\times\vec E= -\frac{\partial\vec B}{\partial t}\nonumber
\end{equation}
\cite[p. 17.3; italics mine]{feyn2}.
\end{quote}
Feynman holds that the ``flux rule'' can be applied only when the material to which it is applied does not change. This condition encloses -- ex post --   the case in which the  ``flux rule'' makes wrong predictions (Blondel's experiment) and those that need an  {\em ad hoc} choice of the integration path. However, the basic reason appears to be more profound and due to the fact that, as shown in section \ref{regola}, the ``flux rule'' -- if physically interpreted --  implies physical interactions with velocity higher than that of light. Therefore, it can not be a physical law.
These Feynman's pages are quoted and discussed in many (all?) of the following papers dealing with electromagnetic induction: see, just for a superficial check,  two papers separated by more than forty years \cite{scanlon, mayer}.
\par
Clearly, Feynman and co - authors were unaware of Maxwell's  general law  nor did they knew about Blondel's experiment. Furthermore, they did not fully develop the consequences of their last sentence by writing down the  definition of the induced {\em emf}  as the line integral of the Lorentz force on a unit positive charge. Nonetheless, in \cite[p. 138]{morton}, this definition is attributed to them.
\par
Feynman's reflections stimulated directly or indirectly a series of papers on electromagnetic induction, starting with the one by Scanlon et al \cite{scanlon}. This paper is particularly interesting owing to its central thesis, taken up also  by others \cite{munley}. The thesis is that there are no ``exceptions to the flux rule'' if only one chooses adequately the integration line. The search of an integration path that saves the ``flux rule'' is clearly an {\em ad hoc} choice that intrinsically recognizes that we are dealing with a calculation tool and not with a physical law. In fact, a physical law must be valid for {\em any} integration path, as (\ref{natural}) does. The authors are aware of the fact that one should take into account also the drift velocity of the charge, but they dismiss it on  account of its smallness \cite[p. 699]{scanlon}. However, if we disregard the drift velocity, we can not deduce -- starting from the definition of induced {\em emf} (\ref{natural}) -- the ``flux rule'' for filiform circuits in the approximation of low velocities (section \ref{regola}). Nor can we deal with electromagnetic induction in extended materials \cite[pp. 3 - 4]{epl}.
\par
 Lev Davidovi\v{c} Landau (1908 - 1968) and Evgeny Mikhailovich Lifshitz (1915 - 1985), in their {\em Theoretical Physics Series}, treated electromagnetic induction in the eight volume ({\em Electrodynamics of Continuous Media}),  in a paragraph dedicated to ``The motion of a conductor in a magnetic field'' \cite{landau_ecm}.  If a conductor is in motion in a magnetic field with velocity $\vec v$, the authors consider the reference frame ``in which the conductor, or
some part of it, is at rest at the instant considered''. Then,  taking into account the transformation equations of the fields,  they write, in the low velocity approximation:
 \begin{equation}\label{ll}
    \mathcal E=  \oint_{l} (\vec E + \vec v \times \vec B)\cdot\vec {dl}
 \end{equation}
 where $(\vec E + \vec v \times \vec B)\approx \vec E'$ and, as usual, the superscript ($'$)  refers to the reference frame of the circuit  in motion with respect to the source of the magnetic field, which is  at rest in the laboratory reference frame \cite[p. 205]{landau_ecm}. It is worth seeing in detail the various passages, omitted in the text.  We start with the statement that the force on a charge at rest is given by:
 \begin{equation}\label{coulombgen}
    \vec F'= q\vec E'
\end{equation}
Consequently, the electromotive force in the reference frame of the circuit is given by:
\begin{equation}\label{L&L}
    \mathcal E'_C= \oint_{l'} \vec E'\cdot\vec {dl'}
\end{equation}
where the subscript $C$ stays for `Coulomb' and reminds the fact that the force exerted on a point charge is the one contemplated
in {\em Electrostatics}.
Then, by using the transformation equations for the fields and the coordinates, we gets:
\begin{equation}\label{L&L2}
\mathcal E'=\Gamma\oint_{l} (\vec E + \vec v \times \vec B)\cdot\vec {dl}=\Gamma \mathcal E\approx \mathcal E
\end{equation}
  At the end of the paragraph,  the authors discuss the unipolar induction.
    They consider the reference frame in which the magnet is at rest, namely a rotating non inertial reference frame. In this frame, the external circuit appears as rotating in the opposite direction. The authors write that the induced {\em emf} is given by:
 \begin{equation}\label{rotation}
    \mathcal E= \oint_{ext}(\vec v\times\vec B)\cdot \vec{dl}=- \oint_{ext} [\vec B \times (\vec r \times \vec \Omega)]\cdot \vec{dl}
 \end{equation}
 where $\vec\Omega$ is the angular velocity of the magnet in the laboratory reference frame and $\vec r$ the distance of a point of the external circuit from magnet axis \cite[p. 209]{landau_ecm}.
Then, the induced {\em emf} should be located {\em somewhere} in the external circuit and not along a radius of the magnet, as predicted by using the inertial reference frame of the laboratory (figure \ref{unipolar}).   Furthermore, the induced {\em emf} should depend on the form of the external circuit.
 Instead, the experiments carried out by M\"{u}ller \cite{muller} confirm that the induced {\em emf} is localized in a radius of the magnet and its value is given by  equation (\ref{fem_faraday}) or its variation (\ref{radius}).
\par
In {\em Classical Electrodynamics} by John David Jackson (1925 - 2016) \cite{jackson}, the electromagnetic induction is dealt with in a   section entitled ``Faraday's Law of Induction''  and the treatment is the same as that of the book's first edition \cite[pp. 170 -173]{jack1}.  After having summarized Faraday's principal experiments, it is stated that ``Faraday attributed the transient current flow to a changing
magnetic flux linked by the circuit'' \cite[p. 208]{jackson}. This historical falsehood is common to many texts and is accompanied by the improper  denotation of the ``flux rule'' as the ``Faraday's law''.
 Then,  ``Faraday's law'' is written as:
\begin{equation}\label{flussojack}
    \mathcal E = - k \frac{d\Phi}{dt}
\end{equation}
where $k$ is a constant depending only on the choice of the units for the electric and magnetic quantities ($k=1$ in the SI units System). Jackson adopts the definition of induced {\em emf} (\ref{emiC}); as shown in section \ref{nota}, this definition, if applied  in the laboratory reference frame to moving circuits,  together with equation (\ref{identity}), leads to the wrong prediction of equation (\ref{noflux}).   In fact, by applying
 equation (\ref{identity}), one gets:
\begin{equation}\label{flussojack2}
    \mathcal E= -k\frac{d\Phi}{dt}=- k \int_S\frac{\partial \vec B}{\partial t}\cdot \hat n\, dS+k\oint_l(\vec v\times\vec B)\cdot\vec{dl}
\end{equation}
Since it is assumed that:
\begin{equation}\label{ovvio}
\mathcal E=\oint_l \vec E\cdot\vec{dl}
\end{equation}
we have:
\begin{equation}\label{flussojack3}
    \oint_l [\vec E - k (\vec v\times\vec B)]\cdot \vec {dl}=-k \int_S\frac{\partial \vec B}{\partial t}\cdot \hat n\, dS
\end{equation}
Let us further assume that  ``Faraday's law'' is valid in any inertial reference frame, that the Galilean coordinates transformations are valid and that $\vec B'=\vec B$ \cite[p. 209]{jackson}.
Then, in the reference frame co - moving with the circuit, we can write:
\begin{equation}\label{flussojack4}
    \oint_{l'} \vec E'\cdot \vec {dl'}=-k \int_{S'}\frac{\partial \vec B'}{\partial t'}\cdot \hat n\, dS'=-k \int_S\frac{\partial \vec B}{\partial t}\cdot \hat n\, dS
\end{equation}
Therefore, by equating the first members of equations (\ref{flussojack3}) and (\ref{flussojack4}):
\begin{equation}\label{wrong}
    \vec E'=\vec E - k (\vec v\times\vec B)
\end{equation}
This approximated relation between the values of the electric field in the two reference frames is wrong.
However, instead of (\ref{wrong}), Jackson writes the correct approximated equation $  \vec E'=\vec E + k (\vec v\times\vec B)$, owing to an odd interchange of the vectors  $\vec E$ and $\vec E'$. This interchange is justified by saying that: a) in equation (\ref{flussojack3}) the field is $\vec E'$ and not $\vec E$ since ``\dots it is that field that causes current to flow if a circuit is actually present''; and b) in equation (\ref{flussojack4}) the field is $\vec E$ and not $\vec E'$ since ``Since we can think of the circuit C and surface S as instantaneously at a certain position in space in the laboratory'' \cite[p. 210]{jackson}.
Hence, it is worth developing an alternative derivation based on the  definition of the induced {\em emf}  as the line integral of the Lorentz force on a unit positive charge. As shown in section \ref{mle}, the ``flux rule'' can be derived in the approximation of low velocities for filiform circuits (in which the drift velocity is parallel to any elementary part $\vec{dl}$ of the circuit):
\begin{equation}\label{fluxj}
    \mathcal E=\oint_l [\vec E+(\vec v\times\vec B)]\cdot \vec{dl}= -\frac{d\Phi}{dt}
\end{equation}
Under Jackson's Galilean approximation we can write, in the reference frame co - moving with the circuit:
\begin{equation}\label{fluxjc}
    \mathcal E'=\oint_{l'} \vec E'\cdot \vec{dl'}= -\frac{d\Phi'}{dt'}=-\frac{d\Phi}{dt}
\end{equation}
Hence, since $\vec {dl'}= \vec {dl}$:
\begin{equation}\label{cosi}
    \vec E'=\vec E+\vec v\times\vec B
\end{equation}
Jackson's Galilean approximation made at the beginning of the calculation reminds us that
 Le Bellac and L\'evy - Leblond have shown that, indeed, there are two Galilean limits of Electromagnetism \cite{levy}.\label{levy}
These limits come out by letting $c=1(\sqrt{\varepsilon_0\mu_0})\rightarrow\infty$. This can be done by keeping $\varepsilon_0$ and eliminating $\mu_0=\varepsilon_0/c^2$ by letting $c\rightarrow\infty$ (electric limit); or, by  keeping $\mu_0$ and eliminating $\varepsilon_0=\mu_0/c^2$ by letting $c\rightarrow\infty$ (magnetic limit). More physically, in the electric limit $|\rho|c\gg |j|$, while, in the magnetic limit   $|\rho|c\ll |j|$. Jackson's approximation, if  developed by starting with the  definition of the induced {\em emf} as the line integral of the Lorentz force on a unit positive charge, falls in the magnetic limit.
Recently, this topic has been resumed and expanded by several authors \cite{holland, rouss1, rouss, heras}.
\par
In Wolfgang  Panofsky (1919 - 2007) and Melba Phillips (1907 - 2004) {\em Classical Electricity and Magnetism}, ``Faraday's law'' of induction is treated in chapter IX, entitled ``Maxwell's equations'' \cite[p. 158]{pp}. This treatment is the same as that contained in the first edition of the book  \cite[142 - 146]{pp29}.
 Considered a circuit of resistance $R$ with an {\em emf} $\mathcal E$, ``Faraday's law'' is stated by the equation:
 \begin{equation}\label{emipp}
   IR- \mathcal E=-\frac{d\Phi}{dt}
 \end{equation}
``This means that the current in the circuit differs from that predicted by
Ohm's law by an amount which can be attributed to an additional electromotive
force equal to the negative time rate of change of flux through the
circuit. Note that equation (\ref{emipp}) is an {\em independent experimental law} and is in
no way derivable from any of the relations that have been previously
 used \cite[p. 158; italics mine]{pp}.''
From this experimental law, putting $-d\Phi/dt=\oint_l \vec E\cdot\vec {dl}$ and arguing that this relation holds also in vacuum, Maxwell's equation for the {\em curl} of the electric field is derived.
\par
The discussion of the case of a moving circuit is treated as in Jackson's textbook, namely by oddly interchanging the values of the electric fields in the two reference
frames \cite[pp. 160 - 163]{pp}.
\par
The book {\em Electricity and Magnetism} by Edward Purcell (1912 - 1997) and David Morin \cite{purmor}   is the third edition of the renowned book by Purcell, published  after his death. As in the cases of Jackson's and of Panowski and Phillips' texts, the treatment of electromagnetic induction is the same as in the previous editions.  Purcell's treatment is founded on the idea that physical laws should be derived from experiment. Therefore, the ``flux rule'' -- Faraday's law in Purcell's denotation -- is ``derived'' step by step from three types of thought experiments, in principle reproducible in laboratory. The first is  that of a conducting bar moving in an uniform and constant magnetic field. This thought experiment is treated by using the magnetic component of the Lorentz force and it is described in both reference frames: the laboratory's and the frame co - moving with the bar. The fields transformations, previously derived -- they too, within thought experiments -- are used in the low velocities approximation \cite[pp. 345 - 346]{purmor}.
\par
Then the motion of a rectangular loop in a non - uniform magnetic field is considered. In this way, by using again the magnetic component of the Lorentz force, the ``flux rule'' is derived \cite[pp. 346 - 350]{purmor}. However, since this derivation has been  carried out in a particular case, it is necessary to state and prove the ``flux rule'' as a theorem, valid for any {\em constant} magnetic field. This can be done by deriving our equation (\ref{identity}) with $\partial \vec B/\partial t=0$ \cite[pp. 350 - 352]{purmor}. Of course, this can be accomplished by implicitly assuming that the charge velocity coincides with the the velocity of the circuit element that contains the charge.
The more general case, i.e. that in which the  magnetic field is non uniform and changes with time, is treated by considering three thought experiments
 \cite[p. 355 -- 358]{purmor}. Two laboratory's tables, separated by a curtain,   can be moved and are setup in the following way \cite[fig. 7.15, p. 355]{purmor}. On table 1 we find a coil connected to a battery through a variable resistance. On table 2, there is only a conducting loop connected to a galvanometer. If table 2 is moved with velocity $v$ away from table 1, the galvanometer's needle on table 2 deviates. If, reestablished the initial configuration, table 1 is moved away from table 2 at the same speed $v$, the galvanometer deviates in the same way. Now, after having  reestablished the initial configuration, the current in the coil of table 1 is varied by varying the variable resistance in such a way that the magnetic field at the coil 2 varies as in the experiments I and II. This implies that in coil 2   an {\em emf} is induced and this {\em emf} is identical to that induced during experiments I and II. Since  table 2 does not know what happens to table 1, table 2 sees only the {\em same} deflection of the galvanometer in all three experiments. Since table 2 has seen the same variation of the magnetic field in all three experiments, he concludes that they all are described by Faraday's law:
\begin{equation}\label{purcon}
\mathcal E=\oint_l \vec E\cdot\vec{dl}=-\frac{d\Phi}{dt}
\end{equation}
 if the integration line is {\em stationary}, as it is in table 2.
 \par
It is worth noting that what is measured by table 2 is transparently illustrated by the general law (\ref{leggegen}) written for table 2:
 \begin{equation}\label{leggegen'}
    \mathcal E= -\oint_{l}\frac{\partial \vec A}{\partial t}\cdot \vec
{dl}
    +\oint_{l} (\vec v_{c}\times \vec B)\cdot\vec {dl}
\end{equation}
Since the second line integral is null (for filiform circuits), it turns out that, in table 2, what is measured is due to the time variation of the vector potential: no matter if this time variation is due to a variation of the current in table 1 or/and to a relative motion between the two tables.
\par
As for the other textbooks so far discussed, in the text by Corrado Mencuccini and Vittorio Silvestrini   the ``flux rule'' is considered as the ``law of electromagnetic induction'', here denoted as ``Faraday - Neumann law''.  The induced {\em emf} is defined as:
\begin{equation}\label{msem1}
    \mathcal E= \oint_l \vec E_i\cdot\vec{dl}
\end{equation}
where $\vec E_i$ is the ``induced electromotive field''  and its value is given by
 \cite[p. 353]{ms1}:
\begin{equation}\label{msem3}
   \vec E_i=\frac{\vec F}{q}=\vec E+ \vec v\times \vec B
\end{equation}
where $\vec F$ is, of course, the Lorentz force.
Moreover, it is specified that we should write $\vec v=\vec v_l+\vec v_d$; however, since in filiform circuits $\vec v_d$ is always parallel to $\vec {dl}$, we can safely use equation (\ref{msem3}), where $\vec v$ is the velocity of the line element $dl$.
Overall, the treatment of electromagnetic induction follows standard patterns, characterized by the {\em assumption} of the ``flux rule'', followed by a series of {\em illustrations} that consider different experimental setups. The conceptual difference between Mencuccini and Silvestrini treatment and those of other textbooks, is the fundamental specification  that the induced electromotive field is given by equation (\ref{msem3}). This specification implies that all written equations are correct, though their theoretical hierarchic stand is obscured, owing to the missing straightforward application of equations (\ref{msem1}, \ref{msem3}).
\par
The recent  {\em Modern Electrodynamics } by Andrew Zangwill \cite{andy},  is  characterized  by the fact that there are many references, thus recalling that  the matters dealt with in textbooks rely an a large bulk of investigations.
The Introduction to Chapter II, entitled ``The Maxwell Equations'' starts with the statement:
\begin{quote}\small
    All physical phenomena in our Universe derive from four fundamental forces. Gravity binds stars  and creates the tides. The strong force binds baryons and mesons and controls nuclear reactions. The  weak force mediates neutrino interactions and changes the flavor of quarks. The fourth force, the  Coulomb - Lorentz force, animates a particle with charge $q$ and velocity $\vec v$ in the presence of an electric  field $\vec E$ and a magnetic field $\vec B$:
    \begin{equation}\label{lorandy}
        \vec F = q(\vec E + \vec v × B)
    \end{equation}
      The subject we call electromagnetism concerns the origin and behavior of the fields $\vec E(r, t)$ and $\vec B(r, t)$  responsible for the force (\ref{lorandy}) \cite[p. 29]{andy}.
\end{quote}
Notwithstanding the recognition of the basic role of Lorentz force,  the approach to  electromagnetic induction does not start with the definition of the induced {\em emf} (\ref{natural}), but with the integration of Maxwell's equation (\ref{eqmax2}) \cite[p. 462]{andy}. This procedure is similar to that of section \ref{nota}, but with the fundamental difference that the line integral $\oint_l \vec E \cdot \vec {dl}$ is not assumed as the definition of the induced {\em emf} in a circuit. This entails that the following  equations are correct. The correct definition of what Zangwill denotes as ``Faraday's electromotive force'' is resumed soon after within the approximation of low velocities and the introduction of the drift velocity of the charges, getting in this way equation (\ref{quasiflux}) for the induced {\em emf} \cite[p. 463]{andy}. This treatment privileges the basic Maxwell's equation (\ref{eqmax2}) as the starting point. In doing so, it gives up  the straightforwardness of  the path which starts with the definition of the induced {\em emf} as the line integral of the Lorentz force on a unit positive charge.
\par
The textbooks taken into consideration confirm the longstanding   tradition of presenting electromagnetic phenomena by following their historical development. In teaching practices at university level, this approach constitutes an exception.
Newtonian mechanics and thermodynamics are usually presented starting from a set of postulates. Even so more, the axiomatic approach is used for special and general relativity and quantum mechanics, perhaps after some introductory pages recalling the basic historical steps of the matter. Hertz,  in referring to Maxwell's equations, wrote:
\begin{quote}\small
    These statements form, as far as the ether is concerned,
the essential parts of Maxwell's theory. Maxwell arrived at
them by starting with the idea of action - at - a - distance and
attributing to the ether the properties of a highly polarisable
dielectric medium. We can also arrive at them in other
ways. {\em But in no way can a direct proof of these equations be
deduced from experience}. It appears most logical, therefore,
to regard them independently of the way in which they have
been arrived at, to consider them as hypothetical assumptions,
and to let their probability depend upon the very large
number of natural laws which they embrace. If we take up this point of view we can dispense with a number of auxiliary
ideas which render the understanding of Maxwell's theory
. more difficult, partly for no other reason than that they really
possess no meaning if we finally exclude the notion of direct action - at - a - distance \cite[p. 138, italics mine]{ew}.
\end{quote}
In contrast with Hertz's position, the historical approach  implies, intentionally or not,  the idea that physical laws must be induced cumulatively from experiment without need of recurring, soon or later, to their derivation within an axiomatic theory.
This approach is  exposed to the risk, not always avoided in the texts considered, of dragging electrostatic concepts and definitions into the domain of Electromagnetism. The   definition of the induced {\em emf} as $\mathcal E= \oint \vec E\cdot\vec{dl}$ is a crucial example. This definition is legitimate, but with the warning that it can be used only in the reference frame of the induced, {\em rigid and filiform}, circuit. When the circuit is in motion, the force exerted on a unit positive charge is not $\vec E$ but $(\vec E + \vec v_c\times \vec B)$ and, therefore, the induced {\em emf} must be written as $\oint [\vec E + (\vec v_c\times\vec B)]\cdot \vec {dl}$. Moreover, the definition inherited from Electrostatics ignores the drift velocities of the charges: coherently, since in Electrostatics all charges are at rest.
\par
The introduction of the vector potential is not a formal choice but the only way of writing down an  equation for the induced {\em emf} that is local. The violation of this locality condition appears to be the basic reason for the absence of the vector potential in the treatment of electromagnetic induction. The other face of the medal is the attribution of the rank of  physical law to a calculation tool such as the ``flux rule''.
The subsidiary role attributed to the vector potential may have contributed to the oblivion of Maxwell's general law.
This subsidiary role might be also an heritage left by Heaviside and Hertz who, at the end of the Nineteenth century proposed to exclude the  vector potential from the fundamental equations of Maxwell's theory.  With the words of Hertz:
\begin{quote}\small
    In the construction of
the new theory the potential served as a scaffolding; by its introduction the distance forces
which appeared discontinuously at particular point were replaced by magnitudes which at
every point in space were determined only by the condition at the neighbouring points. But
after we have learnt to regard the forces themselves as magnitudes of the latter kind, there is
no object in replacing them by potentials unless a mathematical advantage is thereby gained.
And it does not appear to me that any such advantage is attained by the introduction of the
vector potential in the fundamental equations; furthermore, one would expect to find in these
equations relations between the physical magnitudes which are actually observed, and not
between magnitudes which serve for calculations only \cite[p. 196]{ew}.
\end{quote}
This Hertz's criterion can be  substituted by  the less restrictive one -- enunciated on page \pageref{pq} and derived by another, most general, Hertz's criterion -- for attributing physical meaning to a physical quantity. The elimination of the vector potential impedes the derivation of a general law of electromagnetic induction that satisfies the locality condition. Therefore, the vector potential satisfies  the strong condition of the criterion. Not to mention the weak condition: at the turn between the Nineteenth and the Twentieth centuries, the role of the potentials has been vindicated by Alfred - Marie Liénard (1869 - 1958) \cite{lienard} and Emil Wiechert (1861 - 1928) \cite{wiechert} in dealing with the electromagnetic field produced by a point charge in arbitrary motion (\ref{pointcharge}). Furthermore, after the unveiling of the relativistic nature of Electromagnetism by Einstein, Maxwell's equations written in terms of the potentials have proved to be the more suited for dealing with this fundamental property.
\par
As for the physical meaning of the vector potential, the only one comment is found -- among the texts considered --
      in Feynman's {\em Lectures}. Feynman speaks of the vector potential as a `real field', where for `real field' it is intended ``a mathematical function we use for avoiding the idea of action at a distance
   \cite[p. 15 - 7]{feyn2}.'' However, Feynman, while speaking of the vector potential as a `real field' in connection with Quantum Mechanics (and Quantum Electrodynamics),   denies it   such a feature in classical Electromagnetism. ``In any region where $\vec B=0$ even if $\vec A$ is not zero, such as outside a solenoid, there is no discernible effect of $\vec A$. Therefore, for a long time it was believed that $\vec A$ was not  a `real field'. It turns out, however, that there are phenomena involving quantum mechanics which show that the field $\vec A$ is in fact a ``real'' field in the sense we have defined it \cite[p. 15 - 8]{feyn2}''.
 Feynman's concept of  `real field' corresponds to -- but not coincides with --   our conditions for attributing a physical meaning to a physical quantity.

\par
  With the exception of  Feynman's and Zangwill's texts, all the  textbooks hold that the ``flux rule'' is the law of electromagnetic induction, thus attributing the rank of a physical law to a calculation tool.  The paths followed for holding such a position are various. It can be said that the ``flux rule'' is, explicitly or implicitly, assumed as the law of electromagnetic induction, leaving to the mathematical developments only the task of supporting the starting assumption. No textbook recognizes explicitly that the ``flux rule'' violates the locality condition. Indirectly, only Jackson recognizes it by noticing that the ``flux rule'' obeys a kind of Galilean invariance.
\par
 The  analysis of the textbooks reminds us that the  precious task attributed to them is that of contributing to the transmission of the  acquired knowledge to the new generations. However, the acquired knowledge is written nowhere   and, hopefully, no one will ever be appointed to write  it.

\section{Electromagnetic induction in some recent research papers}\label{recent}
Recently, Chyba and Hand made a proposal for obtaining electrical energy through electromagnetic induction  in a circuit at rest  on the Earth surface \cite{chyba}. The circuit, viewed from an inertial frame  $K$ centered at the Earth's center, is rotating in the Earth's magnetic field. The proposal has been subjected to an experimental verification with a substantially negative result \cite{terraexp}. The proposal has also triggered a discussion about its theoretical  foundation \cite{terraexp, jeener, chybar}.
\par
The basic idea of the proposal consists in the magnetic manipulation of a part of the  circuit in order to unbalance the contribution to the induced {\em emf} by  the Earth's magnetic field, supposed uniform at the position of the circuit.
Chyba and Hand start with the definition of induced {\em emf} as:
\begin{equation}\label{emchyba}
    \mathcal E= \oint_l (\vec E + \vec v \times \vec B)\cdot \vec {dl}
\end{equation}
where it is implicitly assumed that the  charge velocity $\vec v$ coincides with that of the circuit element that contains the
charge \cite[p. 2]{chyba}.
 In \ref{inertial} it is shown that this definition  leads to a law of electromagnetic induction Lorentz invariant for rigid circuits. In particular, considered the relative  inertial motion of a magnet and a circuit, it is shown that $\mathcal E_{circuit}=\Gamma  \mathcal E_{magnet}$, where $\Gamma$ is the time dilation factor and $\mathcal E_{magnet}$ and $\mathcal E_{circuit}$ is the induced {\em emf} in the reference frame of the magnet and circuit, respectively.
 \par
This relation allows to prove in a simple way that the proposed setup can not yield any  electrical power if the Earth's magnetic field is uniform at the position of the circuit.
 \par
Let us consider the case in which the positive direction of the $x$ axis of the Earth centered inertial frame $K$ coincides -- instantaneously -- with the direction of the velocity vector $\vec V$ of the circuit.
Using the Lorentz invariance of the law of electromagnetic induction for rigid circuits, it is convenient and {\em sufficient} to use
the inertial reference frame $K'$ instantaneously co - moving with the  circuit. In this frame, we have:
\begin{equation}\label{leggegenterra}
    \mathcal E'= \oint_A^B [\vec E'+\vec v\,'\times (\vec B'_{Earth}+\vec B'_{Shield})] \cdot\vec {dl'}+\oint_B^A [\vec E'+\vec v\,'\times \vec B'_{Earth}] \cdot\vec {dl'}
\end{equation}
where AB is the magnetically manipulated part of the circuit, $\vec v\,'$ is the velocity of the circuit element $dl'$ and $\vec E'$ is coming from the relativistic transformations of the fields between the two reference frames $K$ and $K'$.
 Since $\vec v\,'\equiv 0$, equation (\ref{leggegenterra}) reduces to:
\begin{equation}\label{terrafinale}
    \mathcal E'= \oint_{l'} \vec E'\cdot \vec {dl'}
\end{equation}
Since, $\vec E'$ has the same value at every point of the circuit (because the Earth's magnetic field is uniform at the position of the circuit), the integral is null and we have $\mathcal E'=0$ and, hence, also $\mathcal E= \mathcal E'/\Gamma =0$.
Equation (\ref{leggegenterra}) shows at glance that the magnetic manipulation of part of the circuit is useless, while equation (\ref{terrafinale}) shows that the {\em emf} induced in the circuit is null.
\par
The proof presented here is conceptually equivalent to that of   Veltkamp  and Wijngaarden \cite{terraexp}, with the  difference of relying uniquely on the general relation between the induced {\em emf} in the reference frames of a magnet and a circuit in relative inertial motion,  without the necessity of developing detailed calculations in both reference frames.
\par
 Veltkamp  and Wijgaarden  stress that their proof does not exclude a possible effect due to an Earth's magnetic field co - rotating with the Earth \cite[p. 6]{terraexp}.  However, in section \ref{unisec}, we have shown that the hypothesis of a magnetic field co - rotating with a solid magnet is incompatible with MLE electromagnetism and is falsified by Faraday's  and M\"{u}ller's experiments (section \ref{faruni}).
 \par
Finally it is worth stressing that in all the papers cited in this section, the correct definition of induced {\em emf} (\ref{natural}) is used.
\section{Conclusions}
The discussion about the theoretical description of electromagnetic induction has periodically rekindled through about two centuries, in spite of the fact that the basic experiments go back to Faraday's {\em Experimental Researches} and notwithstanding that a `general law' has been derived by Maxwell within a Lagrangian description of electrical currents (section \ref{maxind}).
This situation is very peculiar and the present paper tries to understand why.
A modern reformulation of Maxwell's law is based on the definition of the induced electromotive force as the line integral of the Lorentz's force on a unit positive charge:
\begin{equation}\label{genfine}
    \mathcal E= \oint_l [\vec E+(\vec v_c\times \vec B)]\cdot\vec{dl}; \qquad \vec E=- \nabla  \varphi -\frac{\partial \vec A}{\partial t}
\end{equation}
This equation is, formally, the same as Maxwell's, but with the fundamental difference that the velocity  appearing in it is the velocity of the charge and not the velocity of the line element containing it. The general law is a local law: it correlates what happens in the integration line at the instant $t$ to the values of quantities at the points of the line at the same instant $t$. For rigid circuits, it is Lorentz invariant   (\ref{inertial}). If expressed in terms of the magnetic field, it allows -- in the approximation of low velocities -- the derivation of the ``flux rule'', for {\em filiform} circuits. The ``flux rule'' is a calculation tool and not a physical law, because, if physically interpreted, it implies physical interactions with velocities higher than that of light. Not always it predicts the correct result; it does not say where  the induced {\em emf} is localized; it requires ad hoc choices of the integration paths, thus revealing its being a calculation tool that must be handled with care  (section \ref{regola}).
\par
Maxwell's general law has been rapidly forgotten; instead, the ``flux rule'' has deeply taken root. An analysis of university textbooks, spanned over one century, assumed to be representative on the basis of the authors and/or on their popularity or diffusion,  suggests that one reason for this oblivion may be related to the longstanding tradition of presenting electromagnetic phenomena following their historical development and to the connected, implicit or not, epistemological position according to which physical laws must be cumulatively derived from experiment without  need of recurring, soon or later, to an axiomatic formulation of the matter. But the main reason appears to be the lacking recognition that a physical must obey the locality condition.
The violation of this  condition appears to be the basic reason for the absence of the vector potential in the treatment of electromagnetic induction and for the attribution of the rank of a physical law to a calculation tool such as the ``flux rule''.
On the other hand, the rooting of the ``flux rule'' has been certainly favored by its calculation utility: this  practical feature has largely overshadowed its predictive and epistemological weakness.
 \par
 In the first decades of the Twentieth century, it was  common the idea that some electromagnetic induction experiments with rotating cylindrical magnets could be explained also by assuming that the ``lines of magnetic force'' introduced by Faraday  rotate with the magnet (section \ref{why}). This is a surprising hypothesis, if one takes into account the fact that Faraday's experiments, as repeatedly stressed  by him, prove that the ``lines od magnetic force'' do not rotate. More surprisingly, this hypothesis has been resumed recently. It is shown that the hypothesis of rotating ``lines of magnetic force'' is incompatible with Maxwell - Lorentz - Einstein electromagnetism and is falsified by experiment (section \ref{unisec}).
 \vskip5mm
\par\noindent
{\bf Acknowledgements.} I would like to thank Biagio Buonaura for a critical reading of the manuscript and for his valuable suggestions.
\appendix
\section{Lorentz invariance of the general law of electromagnetic induction}\label{inertial}
Let us consider  a  source of magnetic field at rest in the laboratory reference frame.  A filiform, rigid circuit moves with velocity $V$ along the positive direction of the common $x\equiv x'$ axis.
In the circuit's reference frame, the general law (\ref{natural}) assumes the form:
 \begin{equation}
{\mathcal E'} =    \oint{\vec E'\cdot \vec{dl'}}+ \oint(\vec v'_d \times \vec B')\cdot\vec{dl'}
\end{equation}
where $\vec v_d'$ is the drift velocity of the charges.
Since the second integral is null, because in every circuit element the drift velocity is parallel to $\vec {dl'}$, we are left with  the first line integral.
Taking into account the fields transformation:
 \begin{eqnarray}
   E'_x&=& E_x=0\nonumber\\
   E'_y&=& \Gamma[E_y+ (\vec V\times \vec B)_y]\nonumber\\
    E'_z&=& \Gamma [E_z +(\vec V\times \vec B)_z]\nonumber
\end{eqnarray}
and  the coordinates transformations:
\begin{eqnarray}
dx'&=&\Gamma dx\nonumber\\
dy'&=&dy\nonumber\\
 dz'&=&dz\nonumber
\end{eqnarray}
we get:
\begin{eqnarray}\label{invapp}
    \mathcal E'&=& \Gamma\oint E_xdx+\Gamma \oint [E_y+(\vec V\times\vec B)_y]dy+\Gamma \oint [E_z+(\vec V\times\vec B)_z]dz\nonumber\\
&&\\
&=&\Gamma \oint[\vec E+(\vec V\times\vec B)\cdot\vec{dl}]\nonumber
\end{eqnarray}
In the laboratory reference frame, we have:
\begin{equation}\label{appemilab}
    \mathcal E= \oint (\vec E + \vec V\times\vec B)\cdot \vec {dl}
\end{equation}
Hence:
\begin{equation}\label{finalmente}
    \mathcal E'=\Gamma \mathcal E
\end{equation}
   We have thus shown that the phenomenon of electromagnetic induction, involving electric and magnetic fields, must be treated relativistically, as claimed by Einstein. Since the relative velocity  $V\ll c$, we can put $ \Gamma = 1 $
because the predicted value of $ \Gamma $ differs from one by an amount
experimentally not detectable.

\section{Sources and fields}\label{pointcharge}
\begin{figure}[h]
\centering{
    \includegraphics[width=8cm]{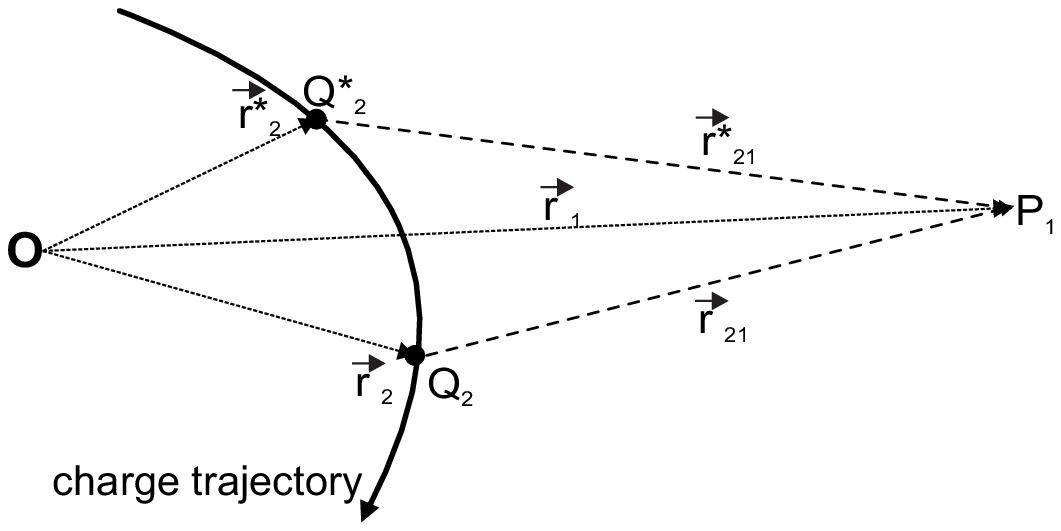}\\
    }
  \caption{Point charge $q$ in arbitrary motion. $P_1$ is the point in which we want to calculate the values of the fields at the instant $t$. $Q_2$ is the position of the charge at the same instant $t$. $Q_2^*$ is the position of the charge at the retarded instant $t^*= t-r^*_{21}/c$.}\label{charge}
\end{figure}
\par\noindent
The electromagnetic potentials produced by a point charge in arbitrary motion, originally and independently studied by  Liénard  \cite{lienard} and Wiechert  \cite{wiechert}, can be written as  (fig. \ref{charge}) \cite[p. 21 - 11]{feyn2}:
\begin{eqnarray}
\varphi (x_1, y_1,z_1,t) & = & {{1} \over {4\pi \varepsilon_0 }}\,
{{q} \over {r\,^*_{21}- {{\vec v\,^*\,\cdot\: \vec
r\,^*_{21}} / {c}} } }\label{scala}\\
&&\nonumber\\
\vec A(x_1,y_1,z_1,t) &= & {{\mu_0} \over {4\pi}}\, {{ q \vec
v\,^*} \over {r\,^*_{21}- \vec v\,^*\, \cdot \, \vec r\,^*_{21} /
c}  } \label{vettor}
\end{eqnarray}
where the asterisk indicates the retarded values of the quantities, i.e. the values  at the retarded instant $t^*= t-r^*_{21}/c$.
By using the   basic relations:
\begin{equation}\label{campipot}
    \vec E= -\nabla \varphi - \frac{\partial \vec A}{\partial t}; \qquad \vec B =\nabla \times \vec A
\end{equation}
one calculates the  fields. The rather laborious calculation yields, for the electric field \cite[pp. 101 - 103, online vers.]{ggib}:
\begin{eqnarray}\label{pointele}
 \vec E&=& {{q}\over{4\pi \varepsilon_0}}{{1}\over{d^{*^{3}}}}
 \left\{\left(1-{{ {v^*}^2}\over{c^2}}\right)
 \left(\vec r\,^*_{21} - r^*_{21} \vec {\frac{v^*}{c}}\right)+\right.
 \nonumber\\
 &+&\left. {{1}\over{c^2}}\vec r\,^*_{21} \times
 \left[\left(\vec r\,^*_{21} - r^*_{21} \vec {\frac{v^*}{c}}
   \right)\times \vec a\,^*  \right]
      \right\}\label{finalmente}
      \end{eqnarray}
      where:
      \begin{equation}\label{dstar}
d^*=r\,^*_{21}- \vec {\frac{v^*}{c}}\cdot \vec r\,^*_{21}
\end{equation}
$d^*$ can be denoted as the reduced retarded distance between the field point and the retarded position of the charge (of course, depending on the sign of the scalar product, this reduced distance can be greater than $r\,^*_{21}$).
      A similar equation is obtained for the magnetic field:
               \begin{eqnarray}\label{magpunti}
                \vec B&=&q\frac{\mu_0}{4\pi}\left\{ \frac{1}{d^{*^{3}}} \left(1 -  \frac{v^{*^{2}}}{c^2} \right) (\vec v^*\times \vec r_{21}^*) +\right.
 \nonumber\\
 &+&\left.\left[\frac{1}{cd^{*^{2}}} (\vec a^*\times \vec r_{21}^*)+\frac{1}{c^2d^{*^{3}}} (\vec a^*\cdot \vec r_{21}^*)(\vec v^*\times \vec r_{21}^*)\right]\right\}
         \end{eqnarray}
         The electric and the magnetic field depend  on the retarded values of the velocity and of the acceleration of  the point charge. They are {\em independently} produced by the charge. These fields propagate with the velocity of light: their values at the point $\vec r_1$ at the instant $t$ depend on the values of the velocity and of the acceleration of the charge at the retarded position  $\vec r_2^*$ and at the retarded instant $t^*=t-r^*_{21}/c$.
      From  these two equations, it can be proved that the two fields are related by the equation:
      \begin{equation}\label{pointmag}
         \vec B = {{1}\over{c}}\hat r^*_{21}\times \vec E
      \end{equation}
    This equation can not be interpreted in any way as a causal relation. It only reflects a property of the two fields as they are {\em independently} produced by the   charge.
\par
This reminds us of an another  subtle question.
Let us consider
  Maxwell's  equation:
\begin{equation}\label{eqmax}
    \nabla \times {\vec E}=-\frac{\partial \vec B}{\partial t}
\end{equation}
It is easy to find  statements according to which this equation shows that a time varying magnetic field {\em causes} an electric field (and symmetrical statements for the equation of the {\em curl} of the magnetic field).  The causes are not  intrinsic properties of mathematical equations, but are superimposed to them by us \cite[pp. 11 - 15]{erq} in agreement with the entire acquired knowledge \footnote{For instance, let us consider the law $\vec F = d\vec p/dt$. As the momentum varies over time, we are inclined to interpret
this equation by saying that the force $ \vec F $ `causes' the variation
of the momentum $ \vec p $. However, there are situations in which the change in momentum `causes'  a force.
Consider, for example, a completely absorbing surface $ S $,
hit by a monochromatic beam of light perpendicular to the surface and directed along the negative direction of the $x$ axis. In this case, we write for the momentum $P_x$ of the surface, if $N$ is the number of photons absorbed in a unit time: $ {dP_x}/{dt}=-N {h\nu}/{c}  =F_x$ and  we can say that the variation of the momentum of photons has
produced a force $ F_x $ on the surface $ S $. A similar situation is found in the kinetic theory of gases: the pressure (force per unit area) on the walls is due the exchange of momentum with the particles.}. Furthermore, the causal chain attributed to an equation must have a correspondence in the experiment. This means that, in the experiment, a causal chain corresponding to that contemplated in the theory must operate. The nowadays  interpretation of Maxwell's equations is that the charges produce (cause) the fields \footnote{It was not so in Nineteenth century. For Hertz, the fields were the primary entities (cause) and the charges the secondary ones (effect). This position is clearly outlined in his  discussion  of the physics  of the capacitor \cite[pp. 20 - 28]{ew}. Hertz's position was not isolated. For instance, in Italy, Galileo Ferraris, shared Hertz's position \cite[pp. 42 - 48, online vers.]{cimento}. Poynting too followed Hertz in this interpretation \cite{poynting}.};  from this point on, no equation can be interpreted by saying that a field (or one its variation)  produces another field. Therefore, equation (\ref{eqmax}) can be read only by saying that it establishes a {\em relation} between the two fields with no causal connections between them.
\par
Incidentally, and in relation to section \ref{marie}, it is worth recalling that in the limit in which the velocity of the charge is small ($v\ll c$) and it varies slowly (so that we can ignore the terms containing the acceleration), we get:
\begin{equation}
    \vec E \approx {{q}\over{4\pi\varepsilon_0r^3_{21}}}\left(\vec r_{21}-r_{21}\frac{\vec v}{c} \right) \label{coulomb_app}
    \end{equation}
 if we ignore the term $v^2/c^2$ with respect to $1$. If $v=0$, we get Coulomb's law.  And, for the magnetic field:
    \begin{equation}
    \vec B \approx  {{\mu_0}\over{4\pi}}\,q {{\vec v \times \vec r_{21}}
    \over{r^3_{21}}}\label{bsbsbs}
\end{equation}
Equation (\ref{bsbsbs}) leads to the Biot - Savart law (\ref{biotsavart}) by using the equation $\vec J=ne\vec v_d$ ($e$ electron charge, $\vec v_d$ drift velocity) and the expression of the Lorentz force.

 \end{document}